\documentclass[12pt]{article}
\usepackage{epsfig, amssymb}
\usepackage{amsmath}
\usepackage{graphicx,epsfig}
\usepackage{color} 
\usepackage{cite}
\usepackage[title]{appendix}
\begin{document}
\thispagestyle{empty}
\begin{flushright}
\end{flushright}

\bigskip

\begin{center}
\noindent{\Large \textbf
{Stochastic quantization and holographic Wilsonian renormalization group of scalar theory with generic mass, self-interaction and multiple trace deformation
}}\\ 
\vspace{2cm} \noindent{Gitae Kim${}^{a}$\footnote{e-mail:kimgitae728@gmail.com }, Ji-seong Chae${}^{a}$\footnote{e-mail:jiseongchae17@gmail.com}, WooCheol Shin${}^{a}$\footnote{e-mail:syc5216@hanyang.ac.kr} and 
Jae-Hyuk Oh${}^{a}$\footnote{e-mail:jaehyukoh@hanyang.ac.kr}}

\vspace{1cm}
  {\it
Department of Physics, Hanyang University, Seoul 04763, Korea${}^{a}$\\
 }
\end{center}

\vspace{0.3cm}
\begin{abstract}
\noindent
We explore the mathematical relationship between holographic Wilsonian renormalization group(HWRG) and stochastic quantization(SQ) of scalar field theory with its generic mass, self-interaction and $n$-multiple-trace deformation on the $d$-dimensional conformal boundary defined in AdS$_{d+1}$ spacetime. We understand that once we define our Euclidean action, $S_E$ as $S_E\equiv -2S_B$, then the stochastic process will reconstruct the holographic Wilsonian renormalization group data via solving Langevin equation and computing stochastic correlation functions. The $S_B$ is given by $S_B=S_{\rm ct}+S_{\rm def}$, where $S_{\rm ct}$ is the boundary counter term and $S_{\rm def}$ is the boundary deformation which gives a boundary condition. In our study, we choose the boundary condition adding (marginal)$n$-multiple trace deformation to the holographic dual field theory.
In this theory, we establish maps between fictitious time, $t$ evolution of stochastic $n$-point, ($2n-2$)-point correlation functions and the (AdS)radial, $r$ evolution of $n$-multiple-trace and ($2n-2$)-multiple-trace deformations respectively once we take identifications of $r=t$ and  between some of constants appearing in both sides. 

\end{abstract}
\newpage
\tableofcontents 

\section{Introduction}
\label{introduction}
It is widely discussed that holography provides a natural energy scale, which is the radial variable ``$r$'' of AdS spacetime, for the $d$-dimensional boundary field theory as a dual of ($d+1$)-dimensional gravity model defined in asymptotically AdS$_{d+1}$ spacetime. One can define a dual field theory on $r=\epsilon$ hypersurface near AdS(conformal)-boundary($r=0$). The generating functional or effective action computed from the dual gravity model shows (UV)divergent terms in it, which are proportional to certain positive integral powers of $1/\epsilon$. To renormalize such terms, one can introduce counter term action. This action contains information of renormalization group flows of effective couplings of deformations to the boundary field theory\cite{Peskin11}. 

The rigorous study of Wilsonian approach of renormalization group to the holographic models is called holographic Wilsonian renormalization group(HWRG)\cite{Heemskerk:2010hk,Faulkner:2010jy} and it is widely studied. 
Especially, studies on holographic renormalization of marginal operators\cite{Witten:2001ua,Aharony:2015afa,Jatkar:2012mm,Oh:2014nfa,Oh:2020zvm} and holographic renormalization in asymptotically Lifshitz spacetime\cite{ Mann:2011hg,Chemissany:2012du} are interesting.
One can think of various gravity models in holography. For free theories,
one can turn on non-normalizable mode of solutions being obtained from equation of motion of the gravity model. The excitation corresponds to turning on single trace deformation to the holographic dual field theory but it turns out that double trace deformation is necessarily appeared in such a case. More interesting case is the gravity models with interactions. Once the interactions are turned on, one needs to consider multiple trace deformations to the boundary field theories and can observe the holographic renormalization group flows of the deformations.
The flows of the multiple trace deformation couplings in energy scale(or in ``$r$'') is given by Hamilton-Jacobi equation, 
\begin{equation}
\label{Hamilton-Jacobi equation,}
\mathcal H_{H}\psi_H=\frac{\partial \psi_H}{\partial \epsilon},
\end{equation}
where the Hamiltonian, $\mathcal H_{H}$ is obtained by Legendre transformation of the Lagrangian density of the dual gravity model. The wave function $\psi_H$ is given by
\begin{equation}
\psi_H=e^{-S_B(\epsilon)},
\end{equation} 
where $S_B(\epsilon)$ contains radial evolutions of boundary deformations. 


On the other hand, we might have holographic Wilsonian renormalization group flow equation in completely different framework: stochastic process. Stochastic quantization(SQ)\cite{Parisi:1980ys,Damgaard:1987rr} is a quantization methodology by employing a first order differential equation, which is so called Langevin equation,
\begin{equation}
\label{Lag-ge-bin}
\frac{\partial \Phi(x,t)}{\partial t}=-\frac{1}{2}\frac{\delta S_E}{\delta\Phi(x,t)}+\eta(x,t),
\end{equation}
where $t$ is called stochastic time, $S_E$ is the Euclidean action to be quantized defined on $d$-dimensional Euclidean manifold, which is a functional of the field $\Phi$. The $\eta$ is called white Gaussian noise which has the following statistical distribution:
\begin{equation}
\mathcal Z\equiv \int [\mathcal D \eta(x,t)] P(\eta(x,t);t)=\int [\mathcal D \eta(x,t)]\exp\left\{ -\frac{1}{2}\int^t_{t_0}dtd^dx {\ }\eta^2(x,t) \right\},
\end{equation}
where $t_0$ is the initial time and  $P(\eta(x,t);t)$ is called probability distribution, which is Gaussian distribution of the noise field.
This Langevin equation describes stochastic time ``$t$'' evolution of the field $\Phi$, which is 
a kind of relaxation process, once we assume that the Euclidean action is unitary theory(positive definiteness of the coefficient of bilinear term in it), interacting with the Gaussian noise.

Such a relaxation process leads the system to be settled down to an equilibrium state as $t$ approach infinity. It turns out that the stochastic correlation function of the field $\Phi(x,t;\eta)$ obtained by solving Langevin equation(\ref{Lag-ge-bin}) precisely gives quantum correlation functions of the Euclidean theory $S_E$ in the equilibrium. We note that for the computation of the correlation functions, the expectation values of the noise field $\eta$ is used, which appears in the solutions of $\Phi(x,t;\eta)$.

This stochastic process also presents Hamiltonian description,
which is given by
\begin{equation}
\label{of the relaxation process, }
\mathcal H_{FP}\psi_S=\frac{\partial \psi_S}{\partial t},
\end{equation}
where the wave function is given by
\begin{equation}
\psi_S=P(\Phi,t)e^{S_E(\Phi)/2},
\end{equation}
and $\mathcal H_{FP}$ is Fokker-Planck Hamiltonian being given by
\begin{equation}
\mathcal H_{FP}=\frac{1}{2}\left(-\Pi+\frac{1}{2}\frac{\delta S_E}{\delta \Phi(x)}\right)
\left(\Pi+\frac{1}{2}\frac{\delta S_E}{\delta \Phi(x)}\right),
\end{equation}
where $\Pi=\frac{\delta}{\delta \Phi}$ is the canonical momentum of the field $\Phi$.


There were some discussion on the relationship between holography and stochastic quantization. In \cite{Lifschytz:2000bj}, the authors argue that Schwinger-Dyson equation of holographic dual gauge theory are eqilibrium condition in stochastic quantization, which also suggests the relation between AdS radial variable $r$ with the stochastic time $t$. In \cite{Mansi:2009mz}, the authors argue that the partition function of the holographic dual gravity theory may correspond to stochastic partition function.

Moreover, recently, there is an interesting attempt to analyze holographic Wilsonian renormalization group(HWRG) by employing stochastic quantization(SQ) method in some literatures\cite{Oh:2012bx,Jatkar:2013uga,Oh:2013tsa,Oh:2015xva,Moon:2017btx,Oh:2021bxx}.
The authors argue that once we identify the radial variable $r$ with the stochastic time $t$ and the Euclidean action $S_E$ with the boundary action $S_B(\epsilon)$ at $r=\epsilon$, where $\epsilon \ll 1$ and $S_E=-2S_B$, then the multiple trace deformations in holographic renormalization group flow can be completely captured by stochastic correlation functions. This also means that the Hamilton-Jacobi equation(\ref{Hamilton-Jacobi equation,}) can share the same description with the stochastic time evolution with the Fokker-Planck Hamiltonian(\ref{of the relaxation process, }). In the previous work, the authors develop the precise relation between the radial evolution of the double trace deformation and the stochastic 2-point function, which is given by
\begin{align}
\left.\frac{\delta^2S_B}{\delta\Phi(r)^2}\right\rvert^{r=t}_{\Phi=0}=\left\langle \Phi(t)\Phi(t)\right\rangle^{-1}_\textrm{S}-\frac{1}{2}\left.\frac{\delta^2S_E}{\delta\Phi(t)^2}\right|_{\Phi=0},
\end{align}
where $\left\langle \Phi(t)\Phi(t)\right\rangle^{-1}_\textrm{S}$ is the inverse of the stochastic 2-point function. The left hand side of the relation is proportional to double trace deformation in a holographic model.

In this paper, we extend the mathematical relation between HWRG and SQ to interacting scalar theory with arbitrary mass and marginal deformation.
We consider a scalar field theory in AdS$_{d+1}$ with its arbitrary mass and self interaction together with a marginal deformation on AdS boundary. For the self interaction, we consider its interaction Lagrangian density as $\mathcal L_{int}\sim\lambda_{2n-2}\phi^{(2n-2)}$, where $\lambda_{2n-2}$ is the self interaction coupling and $\phi$ is the bulk scalar field. We can take the number $n$ to be integral or half-integral number and $n\geq3$. However, once we allow boundary (marginal) deformations only, as explained in the following, then we take $n$ to be an integral number.
For the boundary deformation, we consider a deformation which will give marginal multiple trace deformation to the holographic dual field theory. In the previous literatures\cite{Witten:2001ua,Aharony:2015afa,Kim:2021rik}, it is widely argued that the boundary condition
\begin{equation}
\label{bcfora0}
a_0\sim \bar\sigma_n\phi_0^{n-1}
\end{equation}
gives multiple trace deformation to the holographic dual field theory, where $\bar \sigma_n$ is the multiple trace coupling, while $a_0$ and $\phi_0$ are the coefficients of normalizable and non-normalizable modes of excitation, respectively, for the scalar field in AdS$_{d+1}$.
We understand that when $\nu=\frac{n-2}{2n}d$, then the multiple trace coupling becomes marginal, where $\nu\equiv \sqrt{M^2+\frac{d^2}{4}}$, and $M$ is the scalar field mass. 

A motivation why we consider such a scalar field coupling and a boundary deformation is that our model becomes a generalization of conformally coupled scalar theory in AdS$_4$. This is $d=n=3$ case. Conformally coupled scalar field theory in AdS$_4$ is conjectured to be a gravity dual of a boundary field theory whose effective field theory is a theory of $\phi_0^3$ interaction with a certain kinetic term in 3-dimensional flat spacetime, where the field $\phi_0= \sum_{i=1}^N \psi_i^\dagger \psi_i$ and $\psi_i$ are the component fields enjoying $O(N)$ rotational symmetry
\cite{deHaro:2006ymc}.


To get such a boundary condition, we add a boundary action $S_B=S_{\rm ct}+S_{\rm def}$ near the conformal boundary. $S_{\rm ct}$ is the boundary counter term, which will remove the divergent pieces from the on-shell action of the bulk scalar field theory and $S_{\rm def}$ will provide a boundary condition for the marginal multiple trace deformation to the dual field theory as given in (\ref{bcfora0}). We study such a boundary condition and request that the boundary condition is satisfied even in the $r=\epsilon$ hypersurface, where $\epsilon$ is small but finite. Up to leading order in the bulk coupling $\lambda_{2n-2}$, $\bar\sigma^2_n$ or $\bar\sigma_{2n-2}$, we get the boundary action $S_B$, which precisely gives the boundary condition for marginal multiple trace deformation, where $\bar \sigma_{2n-2}$ is another another possible marginal multiple trace deformation coupling that we can consider.

Now our claim is that {\bf once one defines an Euclidean action, $S_E\equiv -2S_B$ with such a boundary action, $S_B$, it reproduces radial flows of holographic multiple trace deformations by computing stochastic time evolution as a solution of Langevin equation}. We derive the relations for the radial evolutions of double trace, $n$-multiple trace and $(2n-2)$-multiple trace deformations in general ground. To check the relations rigorously, we consider (boundary directional)momentum zero modes of the scalar field solutions in this note. The relations that we obtained are as follows.
Firstly, the relation between radial flows of holographic double trace deformation and stochastic 2-point function is given by
\begin{align}
\left.\frac{\delta^2S_B}{\delta\Phi(r)^2}\right\rvert^{r=t}_{\Phi=0}=\left\langle \Phi(t)\Phi(t)\right\rangle^{-1}_\textrm{S}-\frac{1}{2}\left.\frac{\delta^2S_E}{\delta\Phi(t)^2}\right|_{\Phi=0},
\end{align}
where $\left\langle \Phi(t)\Phi(t)\right\rangle^{-1}_\textrm{S}$ is the inverse of the stochastic 2-point function.
Secondly, the relation between radial flows of holographic $n$-multiple trace deformation and stochastic $n$-point function is given by
\begin{align}
\left.\frac{\delta^nS_B}{\delta\Phi(r)^n}\right\rvert^{r=t}_{\Phi=0}=-\langle \overbrace{\Phi(t)\cdots\Phi(t)}^{n\ \textrm{fields}} {\rangle}_\textrm{S} \bigg[ \left\langle\Phi(t)\Phi(t)\right\rangle^{-1}_\textrm{S} \bigg]^n -\frac{1}{2}\left.\frac{\delta^nS_E}{\delta\Phi(t)^n}\right|_{\Phi=0},
\end{align}
where $\langle \overbrace{\Phi(t)\cdots\Phi(t)}^{n\ \textrm{fields}} {\rangle}_\textrm{S}$ is the stochastic $n$-point function.
Finally, the relation between radial flows of holographic ($2n-2$)-multiple trace deformation and stochastic ($2n-2$)-point function is obtained as
\begin{align}
\left.\frac{\delta^{2n-2}S_B}{\delta\Phi(r)^{2n-2}}\right\rvert^{r=t}_{\Phi=0}
&= -\langle \overbrace{\Phi(t)\cdots\Phi(t)}^{2n-2\ \textrm{fields}} {\rangle}_\textrm{S}\bigg[ \left\langle\Phi(t)\Phi(t)\right\rangle^{-1}_\textrm{S} \bigg]^{2n-2}
\\ \nonumber
&+\frac{(2n-2)!n^2}{2(n!)^2}\bigg[ \left\langle\Phi(t)\Phi(t)\right\rangle^{-1}_\textrm{S} \bigg]^{2n-1} \left\{\langle \overbrace{\Phi(t)\cdots\Phi(t)}^{n\ \textrm{fields}} {\rangle}_\textrm{S}\right\}^2
\\ \nonumber
&-\frac{1}{2}\left.\frac{\delta^{2n-2}S_E}{\delta\Phi(t)^{2n-2}}\right|_{\Phi=0},
\end{align}
where similarly $\langle \overbrace{\Phi(t)\cdots\Phi(t)}^{2n-2\ \textrm{fields}} {\rangle}_\textrm{S}$ is the stochastic ($2n-2$)-point function. 

To test these relations, we need to compute the Euclidean action $S_E$. Once we employ the suggested relation $S_E=-2S_B$, the Euclidean action is given by
\begin{align}
S_E &= \frac{-\frac{1}{2}+\nu}{t} \Phi^2 - 2\bar{\sigma}_n t^{n(-\frac{1}{2}+\nu)} \Phi^n - \left[ 2\bar{\sigma}_{2n-2} + \left(\frac{\bar{\lambda}_{2n-2}}{n-1}-n^2\bar{\sigma}_n^2\right)\frac{t^{2\nu}}{2\nu} \right] t^{2(n-1)(-\frac{1}{2}+\nu)}\Phi^{2n-2}
\\ \nonumber
&+\mathcal O(\bar{\sigma}_n^3, \bar{\sigma}^2_{2n-2} {\rm \ or\ }  \bar{\lambda}^2_{2n-2}),
\end{align}
where $\bar{\sigma}_{2n-2}$ is the $(2n-2)$-multiple trace deformation coupling on the conformal boundary. {\bf We compute stochastic $2$-point, $n$-point and ($2n-2$)-point correlation functions. They are compared with the holographic data, which are the radial evolutions of double, $n$-multiple and ($2n-2$)-multiple trace deformations, respectively. We check that they are perfectly matched with one another via the relations suggested above.}
We note that the field $\Phi$ is the field defined in a new field frame, through a field redefinition as $\Phi=r^{\frac{d-1}{2}}\phi$. We will explain this in Sec.\ref{Another field frame}.

\section{Holographic Model}
\subsection{Holographic model}
We start with the following holographic model of a scalar theory in Euclidean AdS$_{d+1}$ spacetime, which is given by
\begin{equation}
   S= S_{bulk}+S^\prime_B=\int d^{d+1}x\sqrt{g}\left[\frac{1}{2}g^{\mu\nu}\partial_{\mu}\phi\partial_{\nu}\phi+\frac{1}{2}M^2\phi(x)^2+\frac{\lambda_{2n-2}}{2n-2}\phi^{2n-2}\right]+{S'}_B,
\end{equation}
where $\phi$ is the scalar field and $M$ is the scalar field mass. The Euclidean AdS$_{d+1}$ metric $g_{\mu\nu}$ is given by
\begin{align}
ds^2 &= g_{\mu\nu}dx^\mu dx^\nu
\\
&=\frac{1}{r^2}\left(dr^2+d\vec{x}^2\right)
\\
&=\frac{1}{r^2}\left(dr^2+\sum_{i=1}^d \sum_{j=1}^d dx^idx^j\delta_{ij}\right),
\end{align}
where $x^1,x^2,\ldots,x^d$ are the boundary directional coordinate on the conformal boundary(AdS boundary) and $r=x^{d+1}$. The metric factors that appear in the theory are given by
\begin{equation}
\sqrt{g}=r^{-d-1}, \quad g^{\mu\nu}=\delta_{\mu\nu}r^2.
\end{equation}

The $S^\prime_B$ is the Boundary term at $\epsilon$-boundary(at $r=\epsilon$), which is given by
\begin{equation}
    {S'}_B=S_{\rm ct}+S_{\rm def}, 
\end{equation}
where $S_{\rm ct}$ is the counter term action and $S_{\rm def}$ is the boundary deformation,
which will provide a boundary condition to the field $\phi$ near the conformal boundary. The conformal boundary is defined at $r=\epsilon$ together with a condition that $\epsilon\rightarrow 0$.

We may define this action in boundary directional momentum space by employing Fourier transform as 
\begin{equation}
  \phi(x,r)=\frac{1}{(2\pi)^{\frac{d}{2}}}\int e^{-ik_ix_i}\phi(\vec{k}, r)d^dk,
\end{equation}
where $k_ix_i=\sum_{i,j=1}^d k_ix_j\delta_{ij}$.
With such a Fourier transform, our holographic model action transforms into
\begin{eqnarray}
\label{mother-action-momeT}
S&=&\int dr\int d^dk \bigg[\frac{1}{2}r^{1-d}\partial_r\phi(k,r)\partial_r\phi(-k,r)+\frac{1}{2}r^{1-d}k^2\phi(k,r)\phi(-k,r)
\\ \nonumber &+&\frac{1}{2}M^2\phi(k,r)\phi(-k,r)r^{-(d+1)}\bigg]
+\frac{\bar{\lambda}_{2n-2}}{2n-2}\int dr \left[\int \prod^n_{i=1}\phi(k_i,r)d^dk_i\right] r^{-(d+1)}\delta^{(d)}\left(\sum^{2n-2}_{i=1}k_i\right)\\ \nonumber
&+&S_B^\prime,
\end{eqnarray}
where the coupling, 
$\bar {\lambda}_{2n-2}$ is defined in this momentum space as
\begin{equation}
 \bar{\lambda}_{2n-2}\equiv\frac{\lambda_{2n-2}}{(2\pi)^{{(n-2)d}}}.
\end{equation}

\subsection{Near boundary expansion and the boundary condition}
Now we develop the equation of motion of the action(\ref{mother-action-momeT}), being given by
\begin{equation}
    0=-\partial_r\left(r^{1-d}\partial_r\phi(k,r)\right)+r^{1-d}k^2\phi(k,r)+M^2r^{-(d+1)}\phi(k,r)
\end{equation}
The near boundary expansion of the solution is 
\begin{equation}
    \phi(r)={\phi_0}r^{\frac{d}{2}-\nu}+{a_0}r^{\frac{d}{2}+\nu},
\end{equation}
where 
\begin{equation}
    \nu=\sqrt{\frac{d^2}{4}+M^2}
\end{equation}
and $a_0$ is the coefficient of normalizable mode and ${\phi_0}$ is the coefficient of non-normalizable mode of the solution. 
Functional variation of the theory action defined in AdS$_{d+1}$ leads to
\begin{equation}
\delta S^{on-shell}_{bulk}=-\delta{\phi_0}\bigg[ \frac{\frac{d}{2}-\nu}{\epsilon^{2\nu}}{\phi_0}+\left(\frac{d}{2}+\nu\right){a_0}\bigg]-\delta a_0\phi_0\left(\frac{d}{2}-\nu\right)-\delta a_0\cdot a_0 \left(\frac{d}{2}+\nu\right)\epsilon ^{2\nu} + \mathcal{O}(\epsilon ^{4\nu})
\end{equation}
To cancel the divergences and assign the boundary condition, we put $S_B$ as
\begin{equation}
\label{S_B_first}
    S_B=S_{\rm ct}+S_{\rm def}=\frac{1}{2}\sqrt{\gamma(\epsilon)}\left(\frac{d}{2}-\nu\right)\phi^2(\epsilon)+\bar{\sigma}_n\sqrt{\gamma(\epsilon)}\phi^n(\epsilon),
\end{equation}
where the $\epsilon$-boundary metric is given by
\begin{equation}
ds^2=\gamma_{ij}(\epsilon)dx^idx^j\equiv\frac{1}{\epsilon^2}\delta_{ij}dx^idx^j=\frac{d\vec{x}^2}{\epsilon^2}.
\end{equation}
$\gamma_{ij}$ is the boundary metric and $\gamma$ is its determinant ($\sqrt{\gamma(r=\epsilon)}=\epsilon^{-d}$). 
Its functional variation gives
\begin{eqnarray}
\label{deltaS_B-variation}
    \delta S^\prime_B=\epsilon^{-2\nu}\left(\frac{d}{2}-\nu\right){\phi_0}\delta{\phi_0}+\left(\frac{d}{2}-\nu\right)\delta{a_0}{\phi_0}+\left(\frac{d}{2}-\nu\right)\delta{\phi_0}{a_0}+\delta a_0 \cdot a_0 \left(\frac{d}{2}-\nu\right) \epsilon ^{2\nu}
    \\ \nonumber +n\bar{\sigma}_n\epsilon^{-d+n\left(\frac{d}{2}-\nu\right)}\left({\phi_0}^{n-1}\delta{\phi_0}\right)+{O}\left(\epsilon^{-d+n\left(\frac{d}{2}-\nu\right)+2\nu}\right)
\end{eqnarray}

Now we consider a case with a condition
\begin{equation}
\label{the condition-nu}
    \nu=\frac{n-2}{2n}d=d\left(\frac{1}{2}-\frac{1}{n}\right).
\end{equation}
In such a case, the first term in the second line in (\ref{deltaS_B-variation}) is finite at $r=\epsilon$ boundary even when $\epsilon\rightarrow 0$
and that is the term contributing the boundary condition. 
Once we request $\delta S=\delta S_{bulk}^{on-shell}+\delta S_B^\prime=0$, some divergent pieces are canceled in it and the terms left over 
%
\footnote{
\begin{equation}
   \nonumber -d+n\left(\frac{d}{2}-\nu\right)+2\nu=2\nu>0 
\end{equation}
becomes sub-leading when the condition (\ref{the condition-nu}) is satisfied.
}
give an interesting boundary condition. The boundary condition is given by
\begin{equation}
    \left(-2\nu{a_0}+n\bar{\sigma}_n{\phi_0}^{n-1}\right)\delta{\phi_0}=0.
\end{equation}
One can choose one of the two different boundary conditions. First, one can take Dirichlet boundary condition,
\begin{equation}
\quad \delta{\phi_0}=0.
\end{equation}
Another boundary condition is 
\begin{equation}
\label{another-Dirichlet-bc}
    {a_0}=\frac{1}{2\nu}n\bar{\sigma}_n{\phi_0}^{n-1}.
\end{equation}
It is widely discussed that the second boundary condition will give (marginal) $n-$multiple trace deformation to the boundary field theory\cite{Witten:2001ua,Aharony:2015afa}.

We note that the boundary condition (\ref{another-Dirichlet-bc}) is a generalization of Neumann boundary condition and it is possible to impose such a boundary condition in a case that alternative quantization is allowed on the boundary field theory.
The condition for alternative quantization for scalar field
\footnote{
The possible cases for the boundary condition which gives multi trace deformation(alternative quantization is possible) are 
\begin{eqnarray}
\nonumber
\\ \nonumber {\rm When\ }n=3, {\rm then\ } d \leq 6,
\\ \nonumber {\rm When\ }n=4, {\rm then\ } d \leq 4,
\\ \nonumber {\rm When\ }n=5, {\rm then\ }d \leq 3,
\\ \nonumber {\rm When\ }n=6, {\rm then\ } d \leq 3,
\\ \nonumber {\rm When\ }n>6,  {\rm then\ }  d \leq 2.
\end{eqnarray}
}
is given by
\begin{equation}
    -\frac{d^2}{4}\leq M^2 \leq  -\frac{d^2}{4}+1,
\end{equation}
and this restricts $\nu$ as
\begin{equation}
0 \leq \nu \leq 1.
\end{equation}

\subsection{More discussion on the boundary condition with the bulk scalar solution up to $\mathcal{O}(\bar{\lambda}_{2n-2})$} 

Now we develop our discussion on the boundary condition at $r=\epsilon$ boundary upto leading order in $\bar\lambda_{2n-2}$. We will see that in this case, we need to consider that $\epsilon$ is small but finite. To apply our boundary condition, we consider the following form of the boundary action:
\begin{align}
S'_B = \frac{1}{2}\sqrt{\gamma(\epsilon)}\left(\frac{d}{2}-\nu\right)\phi^2(\epsilon) +\sqrt{\gamma(\epsilon)}\bar{\sigma}_n\phi^n(\epsilon) + \frac{\sqrt{\gamma(\epsilon)}}{\epsilon^{2\nu}}\bar{\sigma}_{2n-2}\phi^{2n-2}(\epsilon) + \sqrt{\gamma(\epsilon)}\frac{B(\alpha)}{4\nu}\phi^{2n-2}(\epsilon),
\end{align}
where 
\begin{equation}
B(\alpha)=\alpha_1\frac{\bar{\lambda}_{2n-2}}{n-1}-\alpha_2n^2\bar{\sigma}^2_n,
\end{equation}
and $\alpha_1$ and $\alpha_2$ are constants which will be fixed soon. For the further discussion,  we turn off $\bar{\sigma}_{2n-2}=0$ for a moment.

The equation of motion is given by
\begin{align}
-\partial_r\left(r^{1-d}\partial_r\phi(r)\right) + M^2\ r^{-(1+d)}\phi(r) + \bar{\lambda}_{2n-2}\ r^{-(1+d)}\phi^{2n-3}(r) = 0,
\end{align}
once we turn on the coupling $\bar\lambda_{2n-2}$.
The near boundary expansion of the field, $\phi(r)$ upto the first order in $\bar{\lambda}_{2n-2}$ is
\begin{equation}
\phi(r) = \phi_0 r^{\frac{d}{2}-\nu} + a_0 r^{\frac{d}{2}+\nu} + \frac{\bar{\lambda}_{2n-2}}{8\nu^2}\phi_0^{2n-3}r^{(2n-3)(\frac{d}{2}-\nu)}.
\end{equation}
We also request that $\delta S^{on-shell}_{bulk}+\delta S'_B = 0$ at the $r = \epsilon$ boundary. $\delta S'_B$ and $\delta S^{on-shell}_{bulk}$ are given by
\begin{align}
\delta S'_B &= \delta\phi_0\left(\frac{d}{2}-\nu\right)\frac{\phi_0}{\epsilon^{2\nu}} + \left(\delta\phi_0+\delta a_0\phi_0\right)\left(\frac{d}{2}-\nu\right) + \delta a_0\cdot a_0\left(\frac{d}{2}-\nu\right)\epsilon^{2\nu}
\\ \nonumber
&-n\bar{\sigma}_n\delta\phi_0\left(\phi_0+a_0\epsilon^{2\nu}\right)^{n-1} -n\bar{\sigma}_n\delta a_0\epsilon^{2\nu}\left(\phi_0+a_0\epsilon^{2\nu}\right)^{n-1}
\\ \nonumber
&+\frac{\bar{\lambda}_{2n-2}}{8\nu^2}\delta\phi_0\cdot\phi_0^{2n-3}\cdot 2(n-1)\left(\frac{d}{2}-\nu\right)\epsilon^{2\nu}
\\ \nonumber
&+\left(\alpha_1\frac{\bar{\lambda}_{2n-2}}{2\nu}-\alpha_2\frac{n^2(n-1)}{2\nu}\bar{\sigma}_n^2\right)\left[\delta\phi_0\left(\phi_0+a_0\epsilon^{2\nu}\right)^{2n-3}\epsilon^{2\nu}+ \delta a_0\left(\phi_0+a_0\epsilon^{2\nu}\right)^{2n-3}\epsilon^{4\nu}\right],
\end{align}
and
\begin{align}
\delta S_{bulk} &= g^{rr}\sqrt{g}\ \delta\phi\ \partial_r\phi\big\rvert^{r=\epsilon}
\\ \nonumber
&= -\delta\phi_0\left\{\left(\frac{d}{2}-\nu\right)\frac{\phi_0}{\epsilon^{2\nu}} + \left(\frac{d}{2}+\nu\right)a_0 + \frac{\bar{\lambda}_{2n-3}}{4\nu^2}(2n-3)\left(\frac{d}{2}-\nu\right)\phi_0^{2n-3}\epsilon^{2\nu} \right.
\\ \nonumber
&\left.\hphantom{= -\delta\phi_0} + \frac{\bar{\lambda}_{2n-2}}{8\nu^2}(2n-4)\left(\frac{d}{2}+\nu\right)\phi_0^{2n-4}a_0\epsilon^{4\nu} \right\}
\\ \nonumber
&-\delta a_0 \left\{ \left(\frac{d}{2}-\nu\right)\phi_0 + \left(\frac{d}{2}+\nu\right)a_0\epsilon^{2\nu} + \frac{\bar{\lambda}_{2n-2}}{8\nu^2}(2n-3)\left(\frac{d}{2}-\nu\right)\phi_0^{2n-3}\epsilon^{4\nu} \right\}
\end{align}
respectively.
To find the subleading correction of the boundary condition, we assume that 
\begin{equation}
\label{boundary-a0-ansatz}
a_0 =  \frac{n\bar{\sigma}_n}{2\nu}\phi_0^{n-1} + \beta\epsilon^{2\nu}\phi_0^{2n-3} + \mathcal{O}(\epsilon^{4\nu}),
\end{equation}
and $\beta$ is a constant to be fixed. We also have
\begin{equation}
\delta a_0 =  \frac{n(n-1)\bar{\sigma}_n}{2\nu}\phi_0^{n-2}\delta \phi_0 + \beta(2n-3)\epsilon^{2\nu}\phi_0^{2n-4}\delta \phi_0 + \mathcal{O}(\epsilon^{4\nu}),
\end{equation}
followed by (\ref{boundary-a0-ansatz}).
Demanding that  $\delta S_{bulk}+\delta S'_B = 0$ gives
\begin{align}
-2\nu\beta + \frac{n^2\bar{\sigma}^2_n}{2\nu}(n-1)(1-\alpha_2) + \frac{\bar{\lambda}_{2n-2}}{2\nu}(\alpha_1 - 1) = 0.
\end{align}
Therefore, we understand that when $\alpha_1=\alpha_2=1$ together with $\bar\sigma_{2n-2}=0$, we have $\beta = 0$. This means that if we request the boundary condition to add $n$-multiple trace deformation on the $r=\epsilon$ boundary, we need to add a boundary action, which is given by
\begin{align}
S'_B = \frac{1}{2}\sqrt{\gamma(\epsilon)}\left(\frac{d}{2}-\nu\right)\phi^2(\epsilon) +\sqrt{\gamma(\epsilon)}\bar{\sigma}_n\phi^n(\epsilon) + \frac{\sqrt{\gamma(\epsilon)}}{\epsilon^{2\nu}}\bar{\sigma}_{2n-2}\phi^{2n-2}(\epsilon) + \sqrt{\gamma(\epsilon)}\frac{B}{4\nu}\phi^{2n-2}(\epsilon),
\end{align}
where we need to demand that $\bar\sigma_{2n-2}=0$ and
\begin{equation}
B=\frac{\bar{\lambda}_{2n-2}}{n-1}-n^2\bar{\sigma}^2_n.
\end{equation}
If one turns on $\bar\sigma_{2n-2}$, then which will given another ($2n-2$)-multiple trace deformation to the $r=\epsilon$ boundary.

We close this section with a few remarks.
First, we note that we may have more sub-leading corrections in the boundary and bulk couplings to the boundary action $S^\prime_B$ but we truncate the boundary action upto $O(\bar\lambda_{2n-2})$, $O(\bar\sigma_{2n-2})$ and $O(\bar\sigma_{n}^2)$ for our further discussion.  Second, the minimal form of the boundary action is given when there is a special relation between $\bar\lambda_{2n-2}$ and $\bar\sigma_{n}$ such that
\begin{equation}
\bar\sigma_{n}=\pm\frac{ 1}{n}\sqrt{\frac{\bar\lambda_{2n-2}}{n-1}},
\end{equation}
together with all the $\bar \sigma_m=0$ for $m\neq n$. In this case, the boundary action is given by
\begin{equation}
S'_B = \frac{1}{2}\sqrt{\gamma(\epsilon)}\left[\left(\frac{d}{2}-\nu\right)\phi^2(\epsilon) \pm\frac{ 2}{n}\sqrt{\frac{\bar\lambda_{2n-2}}{n-1}}\bar{\sigma}_n\phi^n(\epsilon).\right]
\end{equation}
Finally, the theory is reduced to conformally coupled scalar theory in AdS spacetime. When $d=3$ and $n=3$, the theory becomes conformally coupled scalar in $AdS_4$ with $\phi^4$ interaction. This theory is conjectured that it corresponds to a boundary $O(N)$-vector theory whose effective action is given by $\phi_0^3$ interaction with standard kinetic term, where $\phi_0$ is a norm-square of the $O(N)$ vector fields. When $d=2$ and $n=4$ is also the case of conformally coupled scalar in AdS$_{3}$ with $\phi^6$ interaction. Therefore, our model is a generalization of these cases. When $d=5$ and $n=5/2$, it corresponds to conformally coupled scalar in AdS$_5$ with $\phi^3$ interaction, but in that case, marginal boundary deformation is tricky to put since $n$ is a fractional number.

\subsection{Another field frame}
\label{Another field frame}
We consider the bulk gravity action in a new field frame of $\Phi$. In the few field frame of $\Phi$, the action is given by
\begin{eqnarray}
S&=& S_{\text{bulk} } + {S'}_B(\epsilon)
\\ &=&\frac{1}{2}\int_{r>\epsilon} dr \ d^dp \left[ \partial_r \Phi_p \partial_r\Phi_{-p}+p^2\Phi_p\Phi_{-p}+\frac{1}{r^2}\left(M^2+\frac{d^2-1}{4}\right)\Phi_p\Phi_{-p}\right]
\\ \nonumber &+&\int dr \frac{\lambda_{2n-2}}{2n-2} r^{(n-2)(d-1)-2}\frac{1}{(2\pi)^{d(n-2)}}\int \left(\prod^{2n-2}_{i=1} d^dp_i \Phi_{p_i}(r)\right)\delta^{(d)}\left(\sum^n_{j=1}\vec{p_j}\right)+S_B(\epsilon)
\end{eqnarray}
where $\epsilon$ is an arbitrary cut-off in the radial direction.\footnote{When $d=3, n=3$ and $d=2, n=4$ cases, the theories are defined on $AdS_4$ and $AdS_3$ with $\phi^4$ and $\phi^6$ interactions explicitly, respectively. Especially, for $n=3$, the theory becomes conformally coupled scalar in $AdS_4$ with $\bar{\sigma}_n\phi^3$-deformation. }
We note that this model is obtained from the scalar theory defined in AdS$_{d+1}$ via a field redefinition, which is given by
\begin{equation}
    \phi=r^{\frac{d-1}{2} } \Phi,
\end{equation}
and the new boundary action $S_B$ is given by
\begin{equation}
S_B(\epsilon)={S'}_B(\epsilon) -\frac{d-1}{4\epsilon} \ \Phi^2(r=\epsilon)
\end{equation}
From the form of $S_B^\prime$, we get $S_B$ being given by
\begin{align}
\label{Boundary-action-new-field-frame}
S_B = \frac{\frac{1}{2}-\nu}{2\epsilon} \Phi^2 +\bar{\sigma}_n \epsilon^{n(-\frac{1}{2}+\nu)} \Phi^n + \left[ \bar{\sigma}_{2n-2} + \left(\frac{\bar{\lambda}_{2n-2}}{n-1}-n^2\bar{\sigma}_n^2\right)\frac{\epsilon^{2\nu}}{4\nu} \right] \epsilon^{2(n-1)(-\frac{1}{2}+\nu)}\Phi^{2n-2}.
\end{align}

In this new field frame, the scalar field theory of $\Phi$ is effectively defined in $\mathbb R_+ \times\mathbb M_{d}$, where $\mathbb M_d$ is the manifold of conformal boundary and  $\mathbb R_+$ is the half of the real line on which the radial variable $r$ runs. This frame is somewhat useful for our discussion on stochastic process since the stochastic process is also defined on the same manifold, where stochastic time $t_0<t<\infty$ is lying along $\mathbb R_+$ and the Euclidean action $S_E$ may be defined on a manifold $\mathbb M_d$ at $t=t_0$ boundary of it.

\section{Holographic Wilsonian renormalization group of scalar theory with $\phi^{2n-2}$-interaction together with multi trace deformation}
In this section, we compute holographic Wilsonian renormalization group of scalar theory with generic mass $M$, self-interaction $\mathcal L_{int}\sim\phi^{2n-2}$, and multiple trace deformation on the boundary field theory which is characterized by the boundary condition, $a_0\sim\bar{\sigma}_n\phi_0^{n-1}$, where $a_0$ and $\phi_0$ are the coefficients of the normalizable and non-normalizable modes of solutions of the field $\phi$ respectively. The $\bar\sigma_n$ is the coupling of the multiple trace deformation.

\subsection{Holographic Wilsonian renormalization group equations}
We start with our discussion on holographic Wilsonian renormalization group of the scalar theory in the new field frame of $\Phi$. Therefore, our holographic model action has a form of
\begin{eqnarray}
S&=&\frac{1}{2}\int dr \ d^dp \left[ \partial_r \Phi_p \partial_r\Phi_{-p}+p^2\Phi_p\Phi_{-p}+\frac{1}{r^2}\left(\nu^2-\frac{1}{4}\right)\Phi_p\Phi_{-p}\right]
\\ \nonumber &+&\int dr \frac{\lambda_{2n-2}}{2n-2} r^{(n-2)(d-1)-2}\frac{1}{(2\pi)^{d(n-2)}}\int \left(\prod^{2n-2}_{i=1} d^dp_i \Phi_{p_i}(r)\right)\delta^{(d)}\left(\sum^n_{j=1}\vec{p_j}\right)+S_B.
\end{eqnarray}
Now let us evaluate the holographic Wilsonian renormalization group flow. We define the conjugate momentum of the field $\Phi$ to derive Hamilton-Jacobi equation, which is given by 
\begin{equation}
\Pi_p=
\frac{\partial \mathcal{L}}{\partial(\partial_r\Phi_p)}=\frac{\delta S_B}{\delta \Phi}=\partial_{r} \Phi_{-p}
\end{equation}
Note that the cut-off $\epsilon$ is an arbitrary value, which means that our holographic model action does not depend on $\epsilon$. Therefore, we request that $\frac{dS}{d\epsilon}=0$. Using this fact, we can derive the Hamilton-Jacobi equation, being given by
\begin{eqnarray}
\label{Hamilton-Jacobi equation-SB'}
\partial_\epsilon {S}_B&=&-\frac{1}{2}\int_{r=\epsilon}d^dp\left[\left(\frac{\delta {S'}_B}{\delta \Phi_p}\right)\left(\frac{\delta {S'}_B}{\delta \Phi_{-p}}\right)-\left(\vert p\vert^2+\frac{1}{r^2}\left(\nu^2-\frac{1}{4}\right)\Phi_p\Phi_{-p}\right)\right]
\\ \nonumber&+& \frac{\lambda_{2n-2}}{2n-2}r^{(n-2)(d-1)-2}\frac{1}{(2\pi)^{d(n-2)}}\int \left[\prod^{2n-2}_{i=1}d^dp_i\Phi_{p_i}(r)\right]\delta^{(d)}\left(\sum^{2n-2}_{j=1}\vec{p_j}\right).
\end{eqnarray}

\paragraph{Evaluation of Hamilton-Jacobi equation}Let us solve the equation by assumption that the trial solution is designed to be a weak field expansion of $\Phi$ in momentum space. The ansatz is given by
\begin{eqnarray}
\label{SB'-ansatz}
{S}_B&=&\Lambda_\Phi(\epsilon)+\int d^dk J_\Phi(\epsilon,k)\Phi_{-k}(\epsilon)+\int d^dk D^{(2)}_\Phi(\epsilon,k)\Phi_k(\epsilon)\Phi_{-k}(\epsilon)
\\ \nonumber &+& \sum^\infty_{m=3}\int\left[ \prod^m_{i=1}d^3k_i\Phi_{k_i}(\epsilon)\right]D^{(m)}_{k_1\dots k_m}(\epsilon)\delta^{(d)}\left(\sum^m_{j=1}k_j\right)
\end{eqnarray}
where $\Lambda(\epsilon), J(\epsilon,k)$ and $D^{(n)}(\epsilon,k)$ are couplings of boundary cosmological constant, single and multiple trace operators which depend on the cut-off $\epsilon$ and boundary momentum $k$.  

By substituting the ansatz(\ref{SB'-ansatz}) into the Hamilton-Jacobi equation(\ref{Hamilton-Jacobi equation-SB'}), we can get the solutions of $\Lambda(\epsilon), J(\epsilon,k)$ and $D^{(n)}(\epsilon,k)$ by requesting that the Hamilton-Jacobi equation is an identical equation in the field $\Phi$. More precisely, the solutions are obtained by comparing the coefficients of $m$-multiple products of field $\Phi$ in the left-hand side and the right-hand side of the equation (\ref{Hamilton-Jacobi equation-SB'}). Then, we obtain the following set of equations.
\begin{itemize}
\item The coefficient of  the Zeroth order in $\Phi$ is given by
 \begin{equation}
     \partial_\epsilon \Lambda_\Phi(\epsilon)=-\frac{1}{2}J_\Phi(\epsilon,p)J_\Phi(\epsilon,-p)
 \end{equation}
\item The coefficient of the First order in $\Phi$ is given by
 \begin{eqnarray}
\partial_\epsilon J_\Phi(\epsilon,k)&=&-2J_\Phi(\epsilon,k)D^{(2)}_\Phi(k,\epsilon)
 \end{eqnarray}
\item The coefficient of the Second order in $\Phi$ is given by
 \begin{eqnarray}
 \partial_\epsilon D^{(2)}_\Phi(k,\epsilon)&=&-\frac{1}{2}\Bigg[ 4D^{(2)}(k,\epsilon)D^{(2)}(-k,\epsilon)-\left(\vert k\vert^2+\frac{1}{\epsilon^2}\left(\nu^2-\frac{1}{4}\right)\right)\Bigg]
 \\ \nonumber&-&3\int d^dk J_\Phi(\epsilon,k) D^{(2)}_{p,k-p,-k}(\epsilon)
 \end{eqnarray}
 \item  The coefficient of $m$-th order in $\Phi$ for $m \geq 3$ is given by
 \begin{align}
 &\partial_\epsilon D^{(m)}_{k_1,\dots,k_m}(\epsilon)\delta^{(d)}\left(\sum^m_{j=1}k_j\right)
 \\ \nonumber
 &= -\frac{1}{2}\Bigg[2J_\Phi(\epsilon,k)(m+1)D^{(m+1)}_{k_1,\dots,k_m,-k}(\epsilon)\delta^{(d)}\left(\sum^{m}_{s=1} k_s-k\right)
 \\ \nonumber
 &+ 4\left(\sum^m_{i=1}D^{(2)}(-k_i,\epsilon)\right)D^{(m)}_{k_1,\dots,k_{m-1},k_m}(\epsilon)\delta^{(d)}\left(\sum^m_{s=1}k_s\right)
 \\ \nonumber
 &+ (1-\delta_{3,m})\sum^{m-1}_{l=3}l(m-l+2)\,\textrm{Perm}\left\{D^{(l)}_{k_1,\dots,k_{l-1},k }(\epsilon) D^{(m-l+2)}_{k_l,\dots,k_m,-k}(\epsilon)\right\} \delta\left(\sum^m_{s=1}k_s\right)\Bigg]
\\ \nonumber &+ \frac{\lambda_{2n-2}}{(2n-2)(2\pi)^{d(n-2)}}\delta_{m,2n-2}\epsilon^{(n-2)(d-1)-2}\int\left[\prod^{2n-2}_{i=1}d^dk_i\Phi_{k_i}(\epsilon)\right]\delta^{(d)}\left(\sum^{2n-2}_{i=1} k_i\right)
 \end{align}
 \end{itemize}

\subsection{Holographic Wilsonian renormalization group solutions}
\paragraph{Solutions of Hamilton-Jacobi equation: Momentum zero solution}
To discuss the simplest and heuristic case, we consider the zero momentum case. For this, we take $\Phi_k(r) \rightarrow \Phi(r)$. The (zero momentum) bulk action is given by
\begin{eqnarray}
S_{\text{bulk} }&=&\frac{1}{2}\int dr\left[ (\partial_r\Phi)^2+\frac{1}{r^2}\left(\nu^2-\frac{1}{4}\right)\Phi^2\right]
\\ \nonumber &+&\int dr \frac{\lambda_{2n-2}}{(2n-2)(2\pi)^{d(n-2)} } r^{(n-2)(d-1)-2}\Phi^{2n-2}+S_B
\end{eqnarray}
and so the Hamilton-Jacobi equation is given by
\begin{eqnarray}
\label{Hamilton-Jacobi-zero mementum scalar1}
    \partial_\epsilon {S}_B&=&-\frac{1}{2}\left[\left(\frac{\delta S_B}{\delta \Phi}\right)^2-\frac{1}{r^2}\left(\nu^2-\frac{1}{4}\right)\Phi^2\right]+\frac{\lambda_{2n-2}}{2n-2}r^{(n-2)(d-1)-2}\frac{1}{(2\pi)^{d(n-2)} }\Phi^{2n-2}.
\end{eqnarray}
To get the solutions, we plug the ansatz into the Hamilton-Jacobi equation
\begin{equation}
S_B=\sum^\infty_{m=2} \ D^{(m)}(\epsilon) \ {\Phi(\epsilon)}^m,
\end{equation}
where we consider the simplest case such that
\begin{equation}
\Lambda_\Phi=J_\Phi=0.
\end{equation}

We list the equations of $D^{(m)}$ in order . The equation for double trace coupling, $D^{(2)}(\epsilon)$ is given by
\begin{equation}
    \partial_\epsilon D^{(2)}(\epsilon)=-\frac{1}{2}\left[ 4{D^{(2)}(\epsilon)}^2-\frac{1}{\epsilon^2}\left(\nu^2-\frac{1}{4}\right)\right].
\end{equation}
It is well-known that the solution of $D^{(2)}(\epsilon)$ is 
\begin{equation}
\label{D2-double-trace-deformation-solution}
	D^{(2)}(\epsilon)=\frac{1}{2}\partial_\epsilon \log {\Phi(\epsilon)}=\frac{1}{2}\frac{\Phi'(\epsilon)}{\Phi(\epsilon)}
\end{equation}
where 
\begin{equation}
\label{D2-double-trace-deformation-solution-2}
    \Phi(\epsilon)=\Phi_0 \epsilon^{\frac{1}{2}-\nu}+a_0\epsilon^{\frac{1}{2}+\nu}.
\end{equation}
For the later use, we give an explicit form of the solution of double trace deformation as
\begin{equation}
D^{(2)}(\epsilon)=\frac{1}{2} \frac{\left(\frac{1}{2}-\nu\right)}{\epsilon}+\frac{\nu \epsilon^{2\nu-1}}{\epsilon^{2\nu}+\frac{1}{\tilde{a}_0}},
\end{equation}
where
\begin{equation}
\tilde{a}_0\equiv\frac{a_0}{\Phi_0}.
\end{equation}

For the case that $m \geq 3$, the general form of $D^{(m)}$  equation is given by
\begin{eqnarray}
\partial_\epsilon D^{(m)}(\epsilon)&=&-\frac{1}{2}\Bigg[ 4mD^{(2)}(\epsilon)D^{(m)}(\epsilon)
\\ \nonumber &+& \sum^{m-1}_{l=3}l(m-l+2)D^{(l)}(\epsilon)D^{(m-l+2)}(\epsilon)(1-\delta_{3,m})
\\ \nonumber &+& \frac{\lambda_{2n-2} \ \delta_{m,2n-2}}{(2n-2)(2\pi)^{d(n-2)}} \ \epsilon^{(n-2)(d-1)-2}\Bigg].
\end{eqnarray}
We note that we request the boundary deformation at $r=\epsilon$ as $\epsilon \rightarrow 0$ which is given by
\begin{equation}
    S_{\text{def}}=\sqrt{\gamma}\epsilon^{\frac{n(d-1)}{2}}\bar{\sigma}_n\Phi_0^n
\end{equation}
where we do not add any other deformations on the conformal boundary except this. More precisely, we consider a case that the boundary deformations $\bar \sigma_m=0$ when $2<m < n$ only, where $\bar{\sigma}_m$ is the coefficient of the $m$ multiple of the field $\Phi$, namely $\Phi^m$ in $S_B$. 
Then $D^{(n)}$ satisfies the boundary condition  $D^{(n)}(\epsilon)\rightarrow\bar{\sigma}_n\epsilon^{-d}\epsilon^{\frac{n(d-1)}{2}}$ at $r=\epsilon$ as $\epsilon \rightarrow 0$.
We also assume that 
$D^{(m)}=0$ for $m \leq n-1$. The condition that $D^{(m)}=0$ for $m \leq n-1$ is not necessarily requested but to make our model simplest, we turn them off, otherwise we have lots of deformation terms and it becomes messy.

As we discussed above, we consider that the non-trivial equation is obtained only for
$m=n$ case when $m>2$. The equation is given by
\begin{eqnarray}
\label{Dn-equation-zero-momentum}
\partial_\epsilon D^{(n)}(\epsilon)&=&-\frac{1}{2}\Bigg[4nD^{(2)}(\epsilon)D^{(n)}(\epsilon)
\\ \nonumber &+&\sum^{n-1}_{l=3}l(n-l+2)D^{(l)}(\epsilon)D^{(n-l+2)}(\epsilon)\left(1-\delta_{3,n}\right)\Bigg].
\end{eqnarray}
Since we suppose $D^{(m)}=0$ for $m \leq n-1$, the terms in the second line, which contain summation in the above equation, vanish.

The solution of the equation(\ref{Dn-equation-zero-momentum}) is given by
\begin{equation}
     D^{(n)}(\epsilon)=\frac{C_n}{\left\{\Phi(\epsilon)\right\}^n}=\frac{C_n}{\left(\Phi_0 \epsilon^{\frac{1}{2}-\nu}+a_0\epsilon^{\frac{1}{2}+\nu}\right)^n},
\end{equation}
where $C_n$ is an integration constant. As $\epsilon \rightarrow 0$, $D^{(n)}$ behaves as
\begin{equation}
D^{(n)}(\epsilon\rightarrow0)= \frac{C_n}{\Phi_0^n}\epsilon^{\left(\frac{d-1}{2}\right)n-d},
\end{equation}
where to derive this expression we use a fact that 
\begin{equation}
\nu-\frac{1}{2}=\frac{d-1}{2}-\frac{d}{n}.
\end{equation}
Since the boundary condition for $ D^{(n)}(\epsilon)$ is given by
\begin{equation}
D^{(n)}(\epsilon\rightarrow0)=  \bar{\sigma}_n\epsilon^{-d+\left(\frac{d-1}{2}\right)n},
\end{equation}
we determine that the constant $C_n$ is to be $\bar{\sigma}_n=\frac{C_n}{\Phi_0^n}$. By introducing a new constant $\tilde{a}_0\equiv\frac{a_0}{\Phi_0}$, we rewrite the solution to be more useful form being given by
\begin{equation}
    D^{(n)}(\epsilon)=\epsilon^{-d+\left(\frac{d-1}{2}\right)n}\bar{\sigma}_n(\epsilon), \quad\text{where}\quad \bar{\sigma}_n(\epsilon)=\frac{\bar{\sigma}_n}{\left(1+\tilde{a}_0\epsilon^{2\nu}\right)^n}.
\end{equation}

As a final step, we obtain the solution of $D^{(2n-2)}(\epsilon)$. The equation for $ D^{(2n-2)}(\epsilon)$ is given by 
\begin{eqnarray}
\partial_\epsilon D^{(2n-2)}(\epsilon) &=& -\frac{1}{2}\Bigg[4(2n-2)D^{(2)}(\epsilon)D^{(2n-2)}(\epsilon)
\\ \nonumber &+& \sum^{2n-3}_{m=3}l(2n-l)D^{(l)}(\epsilon)D^{(2n-l)}(\epsilon)\left(1-\delta_{3,2n-2}\right)\Bigg]
\\ \nonumber &+& \frac{\lambda_{2n-2}}{(2n-2)(2\pi)^{d(n-2)}}\epsilon^{(n-2)(d-1)-2}.
\end{eqnarray}
Since we assume that $D^{(l)}=0$, for $ l \leq n-1$, the form of equation is drastically simplified as
\begin{eqnarray}
\partial_\epsilon D^{(2n-2)}(\epsilon)&=&-\frac{1}{2}\Bigg[4(2n-2)D^{(2)}(\epsilon)D^{(2n-2)}(\epsilon)
\\ \nonumber &+& n^2{D^{(n)}(\epsilon)}^2\Bigg]+\frac{\lambda_{2n-2}}{(2n-2)(2\pi)^{d(n-2)}}\epsilon^{(n-2)(d-1)-2}.
\end{eqnarray}
From this equation, we understand that the $D^{(2n-2)}(\epsilon)$ is affected by $ D^{(2)}(\epsilon)$, $ D^{(n)}(\epsilon)$ and the bulk coupling $\lambda_{2n-2}$ only.
Finally, 
we find a solution for $D^{(2n-2)}(\epsilon)$, which is given by
\begin{eqnarray}
D^{(2n-2)}(\epsilon)&=&\Phi^{-2(n-1)}(\epsilon)\Bigg[ C_{2n-2}+ \int^\epsilon_0 d\bar{\epsilon} \{\Phi(\bar{\epsilon})\}^{2(n-1)}(\bar{\epsilon})^{(n-2)(d-1)-2}
\\ \nonumber &\times& \left\{ \frac{\lambda_{2n-2}}{(2n-2)(2\pi)^{d(n-2)}}-\frac{n^2\bar{\sigma}^2}{2}\left(1+\tilde{a_0}\epsilon^{2\nu}\right)^{-2n}\right\}\Bigg].
\end{eqnarray}
Once we perform the $\int d\bar{\epsilon}$ integration, we finally get 
\begin{align}
\label{LHS of the 2n-2 relation}
D^{(2n-2)}(\epsilon) &= Y_{2n-2}(\epsilon) + \epsilon^{(2\nu-1)(n-1)} \left[ \frac{B}{(2n-1)4\nu} \left\{\frac{1}{\tilde{a}_0}+\epsilon^{2\nu}-\frac{1}{\tilde{a}_0(1+\tilde{a}_0\epsilon^{2\nu})^{2n-2}}\right\} \right.
\\ \nonumber
&\left. +\frac{n^2\sigma_n^2}{4\nu\tilde{a}_0} \left\{\frac{1}{(1+\tilde{a}_0\epsilon^{2\nu})^{2n-1}} - \frac{1}{(1+\tilde{a}_0\epsilon^{2\nu})^{2n-2}} + \frac{1+\tilde{a}_0\epsilon^{2\nu}}{2n-1} - \frac{1}{(2n-1)(1+\tilde{a}_0\epsilon^{2\nu})^{2n-2}} \right\} \right],
\end{align}
where
\begin{align}
Y_{2n-2}(\epsilon) \equiv \frac{\epsilon^{(2\nu-1)(n-1)}}{\left(1+\tilde{a}_0 \epsilon^{2\nu}\right)^{2(n-1)}} \bar{\sigma}_{2n-2} \quad\text{and}\ \ \
\bar{\sigma}_{2n-2} \equiv C_{2n-2} \Phi_0^{-2(n-1)}
\end{align}
together with
\begin{equation}
B \equiv \frac{\bar{\lambda}_{2n-2}}{n-1} - n^2\sigma_n^2.
\end{equation}

\section{Stochastic framework}
\subsection{Brief review of stochastic quantization}

We start with the following Euclidean action,
\begin{equation}
\label{Euclidean-action-position-space}
S_E=\int d^dx \sum_{n=2}^\infty g_n(\nabla^2,t)\Phi^n(x),
\end{equation}
where $\nabla^2\equiv \delta^{ij} \frac{\partial^2}{\partial^i \partial^j}$ is the Laplacian in $d-$dimensional Euclidean space, summed over the spatial indices $i$ and $j$.
By employing the Langevin equation
\begin{equation}
\frac{\partial \Phi(x,t)}{\partial t}=-\frac{1}{2}\frac{\delta S_E}{\delta\Phi(x,t)}+\eta(x,t),
\end{equation}
we obtain the following form of equation from the Euclidean action(\ref{Euclidean-action-position-space}):
\begin{equation}
\frac{\partial \Phi(x,t)}{\partial t}=-\frac{1}{2}\sum_{n=2}^\infty n g_n(\nabla^2,t)\Phi^{n-1}(x)+\eta(x,t).
\end{equation}

The noise field $\eta(x)$ enjoys the Gaussian distribution. Therefore, the noise distribution is called stochastic partition function which is given by
\begin{equation}
\mathcal Z=\int [\mathcal D \eta(x,t)]\exp\left\{ -\frac{1}{2}\int^t_{t_0}dtd^dx {\ }\eta^2(x,t) \right\}.
\end{equation}

Now we define the partition function in momentum space by employing Fourier transform as
\begin{eqnarray}
 \Phi(x,t)=\frac{1}{(2\pi)^\frac{d}{2}}\int e^{-ik_ix_i} \Phi(k,t)d^d k, \\ \nonumber
 \eta(x,t)=\frac{1}{(2\pi)^\frac{d}{2}}\int e^{-ik_ix_i} \eta(k,t)d^d k,
\end{eqnarray}
where $k_ix_i$ is contracted by the Kronecker delta $\delta_{ij}$. In momentum space,
the Euclidean action $S_E$ is given by
\begin{equation}
\label{Euclidean-action-momentum-s}
S_E=\sum_{n=2}^\infty \int \frac{1}{(2\pi)^{n\frac{d}{2}-1}}\left(\prod_{i=1}^n d^d k_i \Phi(k_i,t)\right) g_n(k_1, ... ,k_n ; t) \delta^{(d)}\left(\sum_{j=1}^n k_j\right)
\end{equation}
The Langevin equation in momentum space is given by
\begin{equation}
\frac{\partial \Phi(k,t)}{\partial t}=-\frac{1}{2} \frac{\delta S_E}{\delta \Phi(-k,t)}+\eta(k,t).
\end{equation}
By using the Euclidean action(\ref{Euclidean-action-momentum-s}), one can evaluate the precise form of this Langevin equation in momentum space as
\begin{equation}
\frac{\partial \Phi(k,t)}{\partial t}=-\frac{1}{2} \sum_{n=2}^\infty n\int \left(\prod_{i=1}^{n-1}d^d p_i \Phi(p_i,t)\right)\bar g_n (p_1,...,p_{n-1},-k,t) \delta^{(d)} \left(\sum_{j=1}^{n-1} p_j-k\right) + \eta(k,t),
\end{equation}
where we newly define the kernels $\bar g_n$ as
\begin{equation}
\bar g_n(p_i,t) \equiv \frac{1}{(2\pi)^{\frac{nd}{2}-1}} g_n(p_i,t).
\end{equation}

The stochastic partition function in momentum space is given by
\begin{equation}
\mathcal Z= \int[D\eta(k,t)]\exp \left[ -\frac{1}{2}\int^t_{t_0} dt d^d k{\ }\eta(k,t)\eta(-k,t) \right],
\end{equation}
where $t_0$ is the initial stochastic time. Once we plug the precise form of the Langevin equation into this stochastic partition function to replace the noise field $\eta$ by $\Phi$, we get
\begin{equation}
\mathcal Z= \int[\mathcal D\Phi(k,t)]\exp\{-S_{\rm FP}(\Phi,t) \},
\end{equation}
where $S_{\rm FP}$ is called the Fokker-Planck action. The precise form of the Fokker-Planck action is given by
\begin{eqnarray}
\label{Fokker-Planck-scalar-action1}
\nonumber
S_{\rm FP}&=&-\frac{1}{2}\int^t_{t_0} dt \left[\int d^d k\ \left(\frac{\partial \Phi(k,t)}{\partial t}\right)\left(\frac{\partial \Phi(-k,t)}{\partial t}\right)\right.+\sum_{n=2}^\infty \left(\int \prod_{l=1}^{N} d^d q_l \Phi(q_l,t)\right)\delta^{(d)}\left(\sum_{s=1}^N q_s\right) 
\\ \nonumber
&\times&\left\{\frac{1}{4}\sum_{n=2}^N n(N+2-n) \bar g_n\left(q_1,...,q_n,-\sum_{j=1}^{n-1}q_j ; t\right)
\bar g_{N+2-n}\left(q_n,...q_N,-\sum_{l=n}^N q_l ;t\right) -\frac{\partial \bar g_N (q_1,...,q_n;t)}{\partial t} 
  \right\} 
\\ 
&+&\left.\sum_{m=2}^\infty \frac{\partial}{\partial t} \left\{ \int \left(\prod_{l=1}^m d^d q_l \Phi(q_l;t)\right) \bar g_m(q_1,...,q_m;t)\delta^{(d)} \left(\sum_{s=1}^m q_s\right)\right\}\right].
\end{eqnarray}


{{
\subsection{The relation between stochastic partition function and multiple trace deformations}
Stochastic $n$-point correlation function is given by 
\begin{align}
\left\langle \prod^n_{i=1} \Phi(k_i,t) \right\rangle_\textrm{S}
&= \int \left[\mathcal{D}\Phi(k,t)\right] \mathbf{P}(\Phi(k,t);t) \prod^n_{j=1} \Phi(k_j,t),
\end{align}
where $\mathbf P$ is called probability distribution being given by
\begin{align} 
\mathbf{P}(\Phi(k,t);t) &= e^{-S_p(\Phi(t),t)}
\\ \nonumber
&= \exp\left\{-\sum^{\infty}_{i=2}\left[\prod^i_{j=1}\int d^5k_j\, \Phi(k_j,t) \right] P_n(k_1,\cdots,k_i;t)\,\delta^{(5)}\left(\sum^i_{l=1}k_l\right) \right\}.
\end{align}

As we discussed in Sec.\ref{introduction}, we identify the two different Hamiltonian descriptions of the renormalization group flows: Holographic renormalization group flow and stochastic time evolution with Fokker-Planck Hamiltonian. Their wave functions are also identified as follows. We demand that in the classical limit, by identifying $\psi_H=\exp\left(-S_B\right)$ and $\psi_S=\mathbf{P}(\Phi(k,t);t)\exp\left(\frac{S_E}{2}\right)$, we get a relation
\begin{equation}
S_B = S_p - \frac{1}{2} S_E,
\end{equation}
which is equivalent to
\begin{equation}
\left.D^{(n)}_{k_1,\cdots,k_n}(\epsilon)\right\rvert_{\epsilon=t}
= P_n(k_1,\cdots,k_n;t) - \frac{1}{2} g_n(k_1,\cdots,k_n;t),
\end{equation}
where $g_n$ is given by
\begin{align}
S_E = \sum^\infty_{n=2} \int \left[\prod^n_{i=1}d^5k_i\Phi(k_i,t)\right] g_n(k_1,\cdots,k_n)\delta^{(5)}\left(\sum^n_{j=1}k_j\right).
\end{align}

}}

\subsection{Stochastic partition function}
We start with the general discussion of stochastic process for the theory obtained from the boundary on-shell action. The action contains bi-linear, $n$-multiple and $(2n-2)$-multiple terms of the field $\Phi$ in its Lagrangian density. The explicit form of the partition function is given by
\begin{align}
Z &= \int\left[\mathcal{D}\Phi(k)\right] e^{-S_p}
\\ \nonumber
&= \int\left[\mathcal{D}\Phi(k)\right]\exp\left[-\int P_2(k_1,k_2)\delta^{(d)}(k_1+k_2)\prod^2_{s=1}\Phi(k_s)d^dk_s \right.
\\ \nonumber
&\hphantom{= \int\left[\mathcal{D}\Phi(k)\right]\exp[}- \int P_n(k_1,\cdots,k_n)\delta^{(d)}\left(\sum^n_{i=1}k_i\right)\prod^n_{j=1}\Phi(k_j)d^d\vec{k}_j
\\ \nonumber
&\left. \hphantom{= \int\left[\mathcal{D}\Phi(k)\right]\exp[} -\int P_{2n-2}(k_1,\cdots,k_{2n-2})\delta^{(d)}\left(\sum^{2n-2}_{l=1}k_l\right) \prod^{2n-2}_{m=1}\Phi(k_m)d^d\vec{k}_m +\int J(k)\Phi(k)d^dk\right],
\end{align}
where $P_2(k_1,k_2)$, $ P_n(k_1,\cdots,k_n)$ and $P_{2n-2}(k_1,\cdots,k_{2n-2})$ are the coefficients of bi-linear, $n$-multiple and $(2n-2)$-multiple terms of the field $\Phi$ respectively. We also add a deformation being proportional to an external source $J$ to get a generating functional of the partition function. To evaluate this explicitly, we expand the partition function in terms of $P_n$ as well as $P_{2n-2}$ by assuming that the higher order interaction couplings, $P_n$ and $P_{2n-2}$ are small enough to perform series expansions in them. More precisely, $|P_2|\gg|P_n|\gg|P_{2n-2}|$. Then, the partition function becomes
\begin{align}
\label{Z-manipulated-partition-1}
Z&= \int\left[\mathcal{D}\Phi(k)\right]\exp\left[-\int P_2(k_1,k_2)\delta^{(d)}(k_1+k_2)\prod^2_{s=1}\Phi(k_s)d^dk_s +\int J(k)\Phi(k)d^dk \right]
\\ \nonumber
&\times \left\{ 1 - \int P_n(k_1,\cdots,k_n)\delta^{(d)}\left(\sum^n_{i=1}k_i\right)\prod^n_{j=1}\Phi(k_j)d^d\vec{k_j}\right.
\\ \nonumber
&\hphantom{\times [} +\frac{1}{2!}\int P_n(k_1,\cdots,k_n)\delta^{(d)}\left(\sum^n_{i=1}k_i\right)\prod^n_{j=1}\Phi(k_j)d^d\vec{k_j} \int P_n(k_{n+1},\cdots,k_{2n})\delta^{(d)}\left(\sum^{2n}_{l=n+1}k_l\right)\prod^{2n}_{m=n+1}\Phi(k_m)d^dk_m
\\ \nonumber
&\left.\hphantom{\times [} -\int P_{2n-2}(k_1,\cdots,k_{2n-2})\delta^{(d)}\left(\sum^{2n-2}_{l=1}k_l\right)\prod^{2n-2}_{m=1}\Phi(k_m)d^d\vec{k}_m+... \right\}.
\end{align}
We manipulate the partition function in such a way that we replace every $\Phi$ with $\frac{\delta}{\delta J}$ in the curly bracket in (\ref{Z-manipulated-partition-1}).
After this, we integrate out the field $\Phi$.
Then we get the following form of the generating functional:
\begin{align}
Z&= \left\{ 1 - \int P_n(k_1,\cdots,k_n;t)\delta^{(d)}\left(\sum^n_{i=1}k_i\right)\prod^n_{j=1}\frac{\delta}{\delta J(k_j)}d^d\vec{k}_j \right.
\\ \nonumber
&\hphantom{=[} +\frac{1}{2!}\int P_n(k_1,\cdots,k_n;t) P_n(k_{n+1},\cdots,k_{2n};t) \delta^{(d)}\left(\sum^n_{i=1}k_i\right) \delta^{(d)}\left(\sum^{2n}_{j=n+1}k_j\right) \prod^{2n}_{l=1}\frac{\delta}{\delta J(k_l)}d^d\vec{k}_l
\\ \nonumber
&\left. \hphantom{=[} - \int P_{2n-2}(k_1,\cdots,k_{2n-2};t) \delta^{(d)}\left(\sum^{2n-2}_{l=1}k_l\right) \prod^{2n-2}_{m=1}\frac{\delta}{\delta J(k_m)}d^d\vec{k}_m \right\}
\\ \nonumber
&\times \exp \left[\frac{1}{4} \int d^dp_1 d^dp_2 \frac{\delta^{(d)}(p_1+p_2)}{P_2(p_1,p_2)} J(p_1)J(p_2) \right]
\end{align}

By using the form of the generating functional, we compute 2-point, $n$-point and $(2n-2)$-point correlation functions upto tree level in order.
Tree level 2-point correlation function is given by
\begin{align}
\left\langle \Phi(k_1,t)\Phi(k_2,t) \right\rangle_\textrm{S}
&= \frac{\delta^2 \log Z}{\delta J(k_1)\,\delta J(k_2)} = \frac{1}{2P_2(k_1,k_2;t)} \delta^{(d)}(k_1 + k_2),
\end{align}
and the $n$-point function is
\begin{align}
\left\langle \prod^n_{i=1} \Phi(k_i,t) \right\rangle_\textrm{S}
&= \frac{\delta^n \log Z}{\prod^n_{i=1} \delta J(k_i)} = -n! P_n(k_1,\cdots,k_n;t) \prod^n_{i=1} \frac{1}{2 P_2(k_i,-k_i;t)}\,\delta^{(d)}\left(\sum^n_{j=1}k_j\right).
\end{align}
Finally we compute $(2n-2)$-point correlation function and it is given by
\begin{align}
\left\langle \prod^{2n-2}_{i=1} \Phi(k_i,t) \right\rangle_\textrm{S}
&= \frac{\delta^{2n-2} \log Z}{\prod^{2n-2}_{i=1} \delta J(k_i)}
\\ \nonumber
&= -(2n-2)! P_{2n-2}(k_1,\cdots,k_{2n-2};t) \prod^{2n-2}_{i=1} \frac{1}{2 P_2(k_i,-k_i;t)}\,\delta^{(d)}\left(\sum^{2n-2}_{j=1}k_j\right)
\\ \nonumber
&\hphantom{=} +\frac{1}{2!}\cdot n^2 \frac{1}{2P_2(q,-q;t)} P_n(k_1,\cdots,k_{n-1},q;t)P_n(k_n,\cdots,k_{2n-2},-q;t)
\\ \nonumber
&\hphantom{=} \times (2n-2)!\prod^{2n-2}_{i=1}\frac{1}{2P_2(k_i,-k_i;t)} \delta^{(d)}\left(\sum^{2n-2}_{j=1} k_j\right)
\\ \nonumber
&= (2n-2)!\prod^{2n-2}_{i=1}\frac{1}{2P_2(k_i,-k_i;t)} \delta^{(d)}\left(\sum^{2n-2}_{j=1} k_j\right)
\\ \nonumber
&\hphantom{=} \times \textrm{Perm}\left[ \frac{n^2 P_n(k_1,\cdots,k_{n-1},q;t)P_n(k_n,\cdots,k_{2n-2},-q;t)}{4 P_2(q,-q;t)} - P_{2n-2}(k_1,\cdots,k_{2n-2};t) \right],
\end{align}
where Perm denotes all possible permutations of the momentum labels in the square bracket. Namely,
\begin{equation}
{\rm Perm}\{A(k_1,k_2,...k_m)\}=\frac{1}{m!}\left\{A(k_1,k_2,...k_m)+ {\rm all\ possible\ permutation\ of\ } k_1, k_2,...k_m {\rm\ in\ }A \right\}.
\end{equation}
The inverse relations of the above correlation functions are written as
\begin{align}
P_2(k_1,k_2;t) &= \frac{1}{2} \left\langle \Phi(k_1,t)\Phi(k_2,t) \right\rangle_\textrm{S}^{-1},
\\
P_n(k_1,\cdots,k_n;t) &= -\frac{1}{n!} \left\langle \prod^n_{i=1} \Phi(k_i,t) \right\rangle_\textrm{S} \prod^n_{j=1} \left\langle \Phi(k_j,t)\Phi(-k_j,t) \right\rangle_\textrm{S}^{-1},
\end{align}
and
\begin{align}
&P_{2n-2}(k_1,\cdots,k_{2n-2};t)
\\ \nonumber
&= -\frac{1}{(2n-2)!} \left\langle \prod^{2n-2}_{i=1} \Phi(k_i,t) \right\rangle_\textrm{S} \prod^{2n-2}_{j=1} \left\langle \Phi(k_j,t)\Phi(-k_j,t) \right\rangle_\textrm{S}^{-1}
\\ \nonumber
&+ n^2\,\textrm{Perm}\left[ \frac{P_n(k_1,\cdots,k_{n-1},q;t)P_n(k_n,\cdots,k_{2n-2},-q;t)}{2 P_2(q,-q;t)} \right]
\\ \nonumber
&= -\frac{1}{(2n-2)!} \left\langle \prod^{2n-2}_{i=1} \Phi(k_i,t) \right\rangle_\textrm{S} \prod^{2n-2}_{j=1} \left\langle \Phi(k_j,t)\Phi(-k_j,t) \right\rangle_\textrm{S}^{-1}
\\ \nonumber
&+ \frac{n^2}{(n!)^2} \prod^{2n-2}_{j=1} \left\langle \Phi(k_j,t)\Phi(-k_j,t) \right\rangle_\textrm{S}^{-1} \textrm{Perm}\left[ \left\langle \prod^{n-1}_{l=1} \Phi(k_l,t)\Phi(q,t) \right\rangle_\textrm{S} \frac{\left\langle \Phi(q,t)\Phi(-q,t) \right\rangle_\textrm{S}^{-1}}{2} \left\langle \prod^{2n-2}_{m=n} \Phi(k_m,t)\Phi(-q,t) \right\rangle_\textrm{S} \right].
\end{align}

Finally we apply the relation that we obtained.
By the relation
\begin{align}
\left.D^{(n)}_{k_1,\cdots,k_n}(\epsilon)\right\rvert_{\epsilon=t}
&= P_n(k_1,\cdots,k_n;t)-\frac{1}{2}\,g_n(k_1,\cdots,k_n;t),
\end{align}
we get the following relations.
Firstly, we get the relation between stochastic 2-point function and the radial flow of the double trace deformation, which is given by
\begin{align}
\label{relation-2point-and-radial-flow}
\left.\frac{\delta^2S_B}{\prod^2_{i=1}\delta\Phi(k_i,r)}\right\rvert^{r=t}=\left\langle \Phi(k_1,t)\Phi(k_2,t)\right\rangle^{-1}_\textrm{S}-\frac{1}{2}\frac{\delta^2S_E}{\prod^2_{i=1}\delta\Phi(k_i,t)}.
\end{align}
This relation is obtained and extensively discussed in \cite{Oh:2012bx}.
Secondly, we get the relation between stochastic $n$-point function and $n$-multiple-trace ($n$-multiple of single trace operators) being given by
\begin{align}
\label{relation-npoint-ntrace}
\left.\frac{\delta^nS_B}{\prod^n_{i=1}\delta\Phi(k_i,r)}\right\rvert^{r=t}=-\left\langle \prod^n_{i=1}\Phi(k_i,t)\right\rangle_\textrm{S}\prod^n_{j=1}\left\langle\Phi(k_j,t)\Phi(-k_j,t)\right\rangle^{-1}_\textrm{S}-\frac{1}{2}\frac{\delta^nS_E}{\prod^n_{i=1}\delta\Phi(k_i,t)}.
\end{align}
Finally, the relation between stochastic $(2n-2)$-point function and $(2n-2)$-multiple-trace operator is obtained as
\begin{align}
\label{relation-2n-2point-2n-2trace}
\left.\frac{\delta^{2n-2}S_B}{\prod^{2n-2}_{i=1}\delta\Phi(k_i,r)}\right\rvert^{r=t}&=-\left\langle \prod^{2n-2}_{i=1}\Phi(k_i,t)\right\rangle_\textrm{S}\prod^{2n-2}_{j=1}\left\langle\Phi(k_j,t)\Phi(-k_j,t)\right\rangle^{-1}_\textrm{S}
\\ \nonumber
&+\frac{(2n-2)!n^2}{2(n!)^2}\prod^{2n-2}_{i=1}\left\langle\Phi(k_i,t)\Phi(-k_i,t)\right\rangle^{-1}_\textrm{S}\times\textrm{Perm}\left[\left\langle\left\{\prod^{n-1}_{j=1}\Phi(k_j,t)\right\}\Phi(q,t)\right\rangle_\textrm{S}\right.
\\ \nonumber
&\left.\times\left\langle\Phi(q,t)\Phi(-q,t)\right\rangle^{-1}_\textrm{S}\left\langle\left\{\prod^{2n-2}_{l=n}\Phi(k_l,t)\right\}\Phi(-q,t)\right\rangle_\textrm{S}\right]
\\ \nonumber
&-\frac{1}{2}\frac{\delta^{2n-2}S_E}{\prod^{2n-2}_{i=1}\delta\Phi(k_i,t)}.
\end{align}
The left-hand sides of the above equations denote $\frac{\delta^m S_B}{\prod^m_{i=1}\delta \Phi (k_i,r)} = m!D^{(m)}(k_1,\ldots,k_m;r)$.

We mostly deal with zero momentum case in this note. 
In the zero momentum case, we drop out the arguments $k_i$ in the fields and denote them by $\Phi(t)$. The above relations, \eqref{relation-2point-and-radial-flow}, \eqref{relation-npoint-ntrace}, and \eqref{relation-2n-2point-2n-2trace} are simplified as follows.
The first relation \eqref{relation-2point-and-radial-flow} becomes
\begin{align}
\label{relation-2point-and-radial-flow-zero-momentum}
\left.\frac{\delta^2S_B}{\delta\Phi(r)^2}\right\rvert^{r=t}=\left\langle \Phi(t)\Phi(t)\right\rangle^{-1}_\textrm{S}-\frac{1}{2}\frac{\delta^2S_E}{\delta\Phi(t)^2}.
\end{align}
The second relation \eqref{relation-npoint-ntrace} becomes
\begin{align}
\label{relation-npoint-ntrace-zero-momentum}
\left.\frac{\delta^nS_B}{\delta\Phi(r)^n}\right\rvert^{r=t}=-\langle \overbrace{\Phi(t)\cdots\Phi(t)}^{n\ \textrm{fields}} {\rangle}_\textrm{S} \bigg[ \left\langle\Phi(t)\Phi(t)\right\rangle^{-1}_\textrm{S} \bigg]^n -\frac{1}{2}\frac{\delta^nS_E}{\delta\Phi(t)^n}.
\end{align}
The final relation \eqref{relation-2n-2point-2n-2trace} becomes
\begin{align}
\label{relation-2n-2point-2n-2trace-zero-momentum}
\left.\frac{\delta^{2n-2}S_B}{\delta\Phi(r)^{2n-2}}\right\rvert^{r=t}
&= -\langle \overbrace{\Phi(t)\cdots\Phi(t)}^{2n-2\ \textrm{fields}} {\rangle}_\textrm{S}\bigg[ \left\langle\Phi(t)\Phi(t)\right\rangle^{-1}_\textrm{S} \bigg]^{2n-2}
\\ \nonumber
&+\frac{(2n-2)!n^2}{2(n!)^2}\bigg[ \left\langle\Phi(t)\Phi(t)\right\rangle^{-1}_\textrm{S} \bigg]^{2n-1} \left\{\langle \overbrace{\Phi(t)\cdots\Phi(t)}^{n\ \textrm{fields}} {\rangle}_\textrm{S}\right\}^2
\\ \nonumber
&-\frac{1}{2}\frac{\delta^{2n-2}S_E}{\delta\Phi(t)^{2n-2}}.
\end{align}
The prescription of permutation, Perm$\{...\}$ is dropped out because all the momentum labels are vanished. 

\subsection{Solution of the Langevin equation from the zero momentum Euclidean action}

We start with the Euclidean action that we obtained from the holographic data. We suggest that the Euclidean action for our stochastic process is given by
\begin{equation}
S_E=-2 S_B,
\end{equation}
where $S_B$ is given in (\ref{Boundary-action-new-field-frame}).
Then, the Euclidean action is given by
\begin{align}
\label{Euclidean-action}
S_E = \frac{-\frac{1}{2}+\nu}{t} \Phi^2 - 2\bar{\sigma}_n t^{n(-\frac{1}{2}+\nu)} \Phi^n - \left[ 2\bar{\sigma}_{2n-2} + \left(\frac{\bar{\lambda}_{2n-2}}{n-1}-n^2\bar{\sigma}_n^2\right)\frac{t^{2\nu}}{2\nu} \right] t^{2(n-1)(-\frac{1}{2}+\nu)}\Phi^{2n-2}.
\end{align}
From this action, one can evaluate Langevin Equation, which has a form of
\begin{align}
\label{Phi-Langevin-zero-moo}
\frac{\partial \Phi (t)}{\partial t} &=  \left(\frac{\frac{1}{2}-\nu}{t}\right)\Phi(t) + \eta(t) +n\bar{\sigma}_n t^{n(-\frac{1}{2}+\nu)}\Phi^{n-1}(t)
\\ \nonumber
&+ (n-1)\left[ 2\bar{\sigma}_{2n-2} + \left(\frac{\bar{\lambda}_{2n-2}}{n-1}-n^2\bar{\sigma}_n^2\right)\frac{t^{2\nu}}{2\nu} \right] t^{2(n-1)(-\frac{1}{2}+\nu)}\Phi^{2n-3}(t), \\ \nonumber
&=\left(\frac{\frac{1}{2}-\nu}{t}\right)\Phi(t) +{\rm source\ terms.}
\end{align}
Now let us solve this equation to get the solution of $\Phi(t)$. 
We solve this equation with power expansion order by order in small couplings, $\bar{\sigma}_n$ and $\bar\lambda_{2n-2}$ together with $\bar{\sigma}_{2n-2}$, and will get the solution up to $O(\bar\lambda_{2n-2})$, $O(\bar{\sigma}_{2n-2})$ or $O(\bar{\sigma}_n^2)$. This is consistent with weak field expansion in $\Phi$ too. More precisely, $\Phi(t)\equiv\sum_{l=2}^\infty \Phi_l(t)$ where $|\Phi_l| < |\Phi_{l'}|$ if $l>l'$. 
First, we consider only the first two terms on the right-hand side of \eqref{Phi-Langevin-zero-moo} to get the solution of $\Phi_2$, which are the lowest order terms in the expansion. The solution is given by
\begin{align}
\Phi_2(t) = \int^t_{t_0}dt'\,\frac{t^{\frac{1}{2}-\nu}}{t'^{\frac{1}{2}-\nu}}\,\eta(t'),
\end{align}
where $t_0$ is the initial time, which will be fixed by demanding an appropriate initial condition. 
Next, we consider the sub-leading terms. The first sub-leading term is proportional to $\bar{\sigma}_n$. 
We assume that the solution can be written as
\begin{align}
\Phi_n(t) \equiv  \bar{\sigma}_n f(t).
\end{align}
We plug this trial solution in (\ref{Phi-Langevin-zero-moo}) and get the equation in the first order in $\bar{\sigma}_n$. The equation is given by
\begin{align}
\frac{\partial f(t)}{\partial t} - \left(\frac{\frac{1}{2}-\nu}{t}\right) f(t) = n t^{n(-\frac{1}{2}+\nu)} S_{n-1}(t),
\end{align}
and the solution is given by
\begin{align}
f(t)=n\,t^{\frac{1}{2}-\nu} \int^t_{t'_0}dt{''}\,(t'')^{(n+1)(-\frac{1}{2}+\nu)} \prod_{j=1}^{n-1} \left[ \int^{t''}_{t_0}dt'_j\,\frac{(t'')^{\frac{1}{2}-\nu}}{(t'_j)^{\frac{1}{2}-\nu}}\,\eta(t'_j) \right],
\end{align}
where $t'_0$ is another constant. For simplicity, we denote the product of integrals by
\begin{equation}
S_n(t) \equiv \prod_{j=1}^{n} \left[ \int^t_{t_0}dt'_j\,\frac{t^{\frac{1}{2}-\nu}}{(t'_j)^{\frac{1}{2}-\nu}}\,\eta(t'_j) \right].
\end{equation}
Finally, we consider the equation upto $O(\bar\lambda_{2n-2})$, $O(\bar{\sigma}_{2n-2})$ or $O(\bar{\sigma}_n^2)$ and write its solution as
\begin{align}
\Phi_{2n-2}(t) \equiv g(t).
\end{align}
{
The equation reduces to
\begin{align}
\label{The-equation-reduces-to}
\frac{\partial g(t)}{\partial t} - \frac{\frac{1}{2}-\nu}{t} g(t) =&\,
n^2(n-1)\bar{\sigma}_n^2\,t^{(n-1)(-\frac{1}{2}+\nu)}\left[\int^t_{t_0}dt'\,(t')^{(n+1)(-\frac{1}{2}+\nu)}S_{n-1}(t')\right] S_{n-2}(t)
\\ \nonumber
&+ (n-1)\left[ 2\bar{\sigma}_{2n-2} + \left(\frac{\bar{\lambda}_{2n-2}}{n-1} -n^2\bar{\sigma}_n^2\right)\frac{t^{2\nu}}{2\nu} \right] t^{2(n-1)(-\frac{1}{2}+\nu)} S_{2n-3}(t).
\end{align}
Because its homogeneous solution is given by $g(t)=t^{\frac{1}{2}-\nu}$ if the right-hand side of the equation \eqref{The-equation-reduces-to} vanishes, the solution probably has a form of
\begin{equation}
g(t) \equiv t^{\frac{1}{2}-\nu}\bar{g}(t).
\end{equation}
Then the equation reduces to
\begin{align}
\frac{\partial \bar{g}(t)}{\partial t} =&\, n^2(n-1)\bar{\sigma}_n^2t^{n(-\frac{1}{2}+\nu)}\left[ \int^t_{t_0}dt'\,(t')^{(n+1)(-\frac{1}{2}+\nu)}S_{n-1}(t') \right] S_{n-2}(t)
\\ \nonumber
&+ (n-1)\left[ 2\bar{\sigma}_{2n-2} + \left(\frac{\bar{\lambda}_{2n-2}}{n-1}-n^2\bar{\sigma}_n^2 \right) \right]t^{(2n-1)(-\frac{1}{2}+\nu)}S_{2n-3}(t).
\end{align}
By solving this first order equation in $t$, we get 
\begin{align}
g_n(t) = t^{\frac{1}{2}-\nu}\int^t_{t_0}dt'\Biggr[ & n^2(n-1)\bar{\sigma}_n^2\,(t')^{n(-\frac{1}{2}+\nu)} \left\{ \int^{t'}_{t_0}dt''\,(t'')^{(n+1)(-\frac{1}{2}+\nu)}S_{n-1}(t'') \right\}S_{n-2}(t')
\\ \nonumber
&+ (n-1)\left\{ 2\bar{\sigma}_{2n-2}+\left(\frac{\bar{\lambda}_{2n-2}}{n-1}-n^2\bar{\sigma}_n^2 \right)\frac{(t')^{2\nu}}{2\nu} \right\}(t')^{2(n-1)(-\frac{1}{2}+\nu)} S_{2n-3}(t') \Biggr].
\end{align}
}

In sum, the solution $\Phi(t)$ for \eqref{Phi-Langevin-zero-moo} upto $O(\bar\lambda_{2n-2})$, $O(\bar\sigma_{2n-2})$ or $O(\bar{\sigma}_n^2)$ is given by
\begin{align}
\label{Phi-Langevin-sol-final}
\Phi(t) &= \int^t_{t_0}dt'\,\frac{t^{\frac{1}{2}-\nu}}{t'^{\frac{1}{2}-\nu}}\,\eta(t')
+ n \bar{\sigma}_n t^{\frac{1}{2}-\nu} \int^t_{t'_0}dt'\,(t')^{(n+1)(-\frac{1}{2}+\nu)}S_{n-1}(t')
\\ \nonumber
&+ t^{\frac{1}{2}-\nu}\int^t_{t''_0}dt'\Biggr[ n^2(n-1)\bar{\sigma}_n^2\,(t')^{n(-\frac{1}{2}+\nu)} \left\{ \int^{t'}_{t_0}dt''\,(t'')^{(n+1)(-\frac{1}{2}+\nu)}S_{n-1}(t'') \right\}S_{n-2}(t')
\\ \nonumber &\qquad\qquad\qquad
+ (n-1)\left\{ 2\bar{\sigma}_{2n-2}+\left(\frac{\bar{\lambda}_{2n-2}}{n-1}-n^2\bar{\sigma}_n^2 \right)\frac{(t')^{2\nu}}{2\nu} \right\}(t')^{2(n-1)(-\frac{1}{2}+\nu)} S_{2n-3}(t') \Biggr].
\end{align}
This solution can be interpreted by using the following fragments constituting its Feynman diagrams. We may assign a diagram to each component consisting of the terms of $\Phi(t)$ as follows:
\begin{align}
\frac{t^{\frac{1}{2}-\nu}}{(t')^{\frac{1}{2}-\nu}} &={\rm (propagater)},
\\ \nonumber
\eta(t') &={\rm (noise\ source)},
\\ \nonumber
n\bar{\sigma}_n(t')^{n\left(-\frac{1}{2}+\nu\right)} &={\rm (}n{\rm -point\ vertex)},
\\ \nonumber
n^2(n-1)\bar{\sigma}_n^2\,(t')^{n(-\frac{1}{2}+\nu)} \left\{ \int^{t'}_{t_0}dt''\,(t'')^{(n+1)(-\frac{1}{2}+\nu)}S_{n-1}(t'') \right\}S_{n-2}(t')&={\rm  (}(2n-2){\rm -point\ vertex), 1st\  type},
\\ \nonumber
(n-1)\left[2\bar{\sigma}_{2n-2}+\left(\frac{\bar{\lambda}_{2n-2}}{n-1}-n^2\bar{\sigma}^2_n\right)\frac{(t')^{2\nu}}{2\nu}\right](t')^{2(n-1)(-\frac{1}{2}+\nu)} &={\rm (}(2n-2){\rm -point\ vertex), 2nd\  type}.
\end{align}

\begin{figure}[b!]
\centering
\includegraphics[width=180mm]{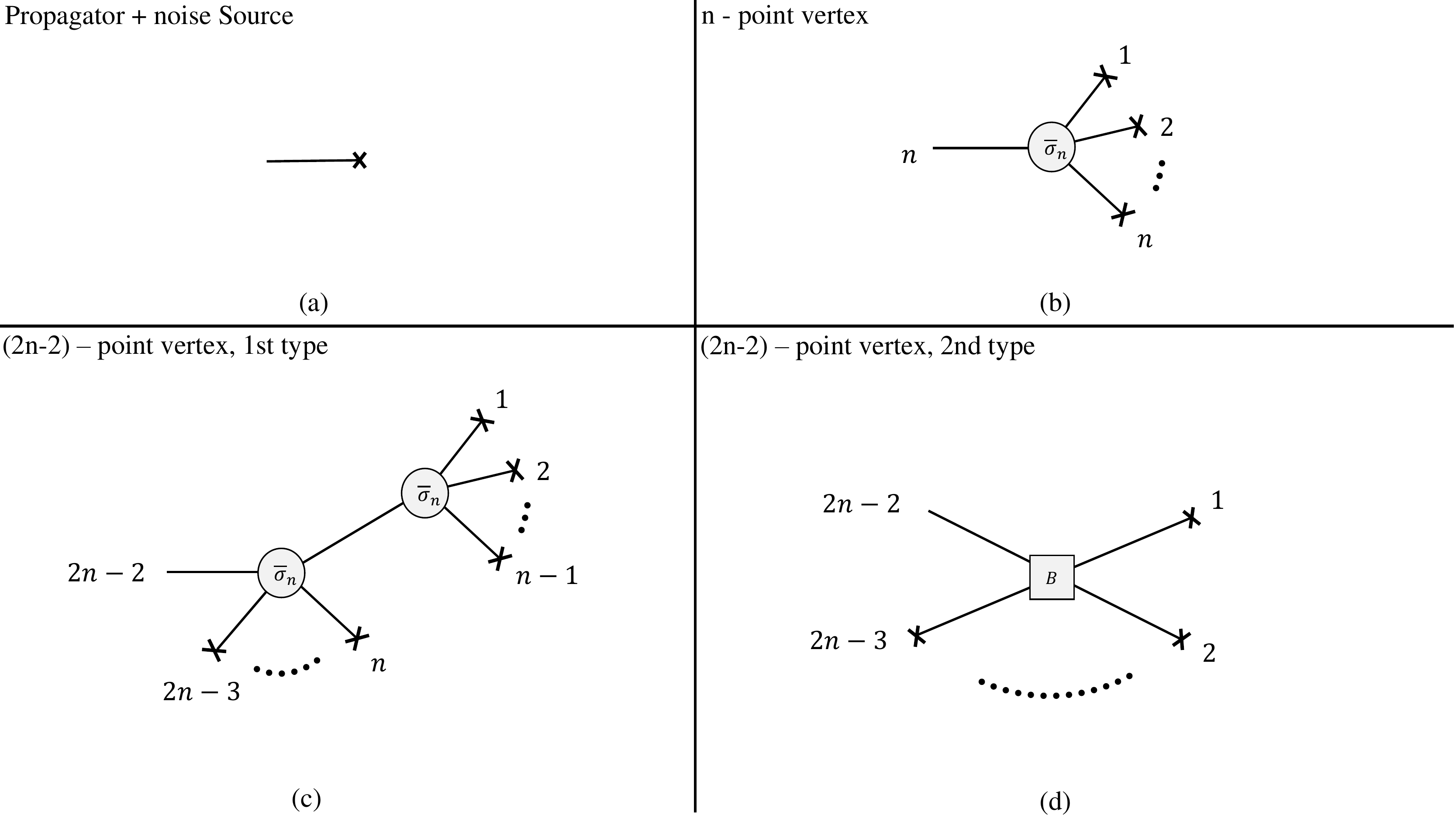}
\caption{Diagrams assigned to each component consisting of $\Phi(t)$. A thin line represents a propagator, and an X mark represents a noise source. A circle represents an $n$-point vertex, and a square represents a $(2n-2)$-point vertex.}
\label{scffig1}
\end{figure}

\subsection{Computations of stochastic correlation functions}
\subsubsection{Stochastic 2-point function}
\begin{figure}[b!]
\centering
\includegraphics[width=180mm]{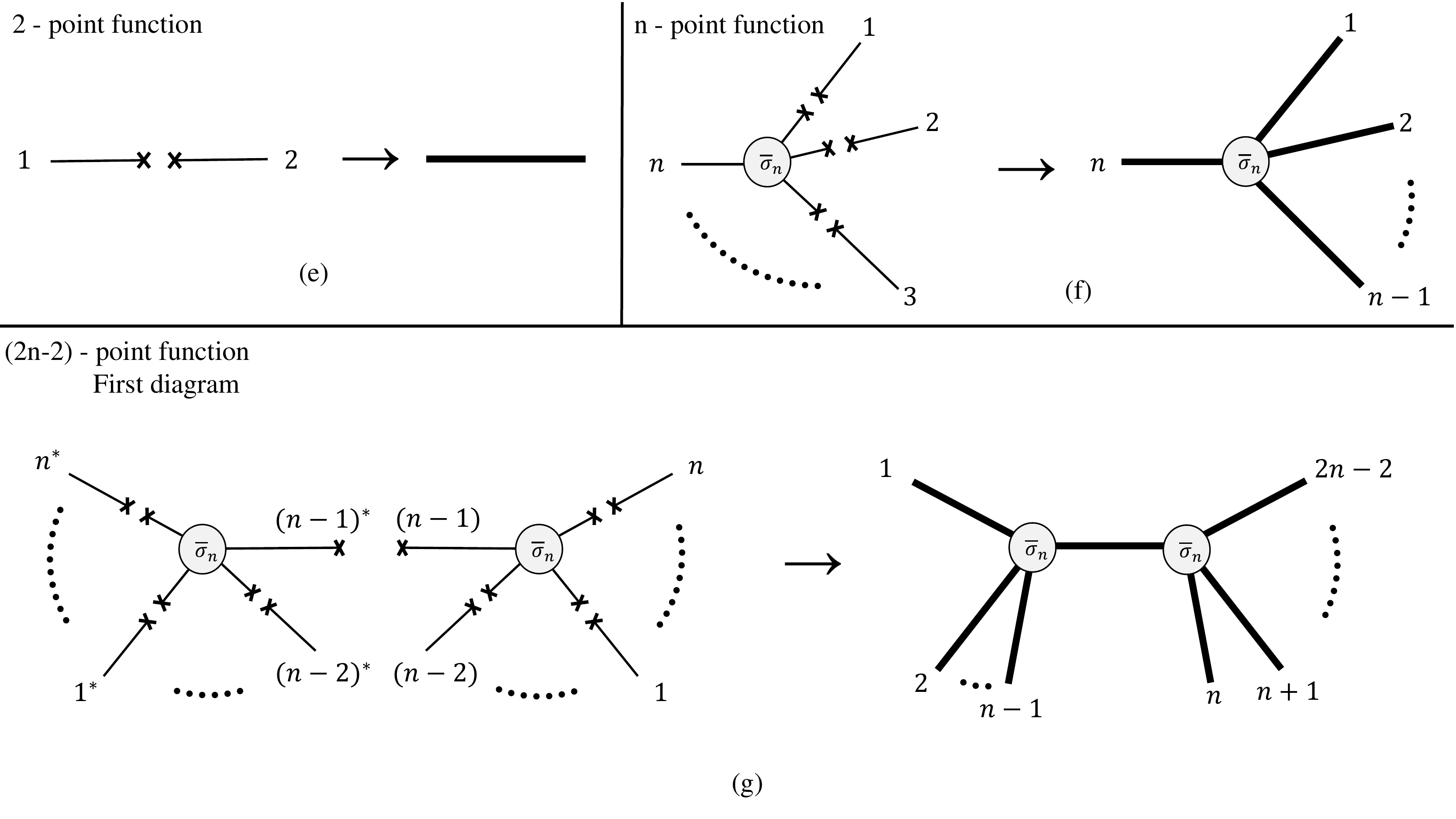}
\caption{Diagrams for the 2-point, $n$-point, and the first type $(2n-2)$-point functions. A noise source, represented by an X mark, is contracted with another noise source. Such a contraction is represented by a thick line.}
\label{scffig2}
\end{figure}

Firstly, we consider the stochastic 2-point function at tree level. The 2-point function is given by
\begin{align}
\label{stochastic-2-point-function-at-tree-level}
\langle \Phi(t)\Phi(t) {\rangle}_\textrm{S} &= \int^t_{t_0} dt' \int^t_{t_0} dt'' \, \frac{t^{\frac{1}{2}-\nu}t^{\frac{1}{2}-\nu}}{(t')^{\frac{1}{2}-\nu}(t'')^{\frac{1}{2}-\nu}} \langle \eta(t')\eta(t'') {\rangle}_\textrm{S}
\\ \nonumber
&= t^{1-2\nu} \int^t_{t_0} dt' \int^t_{t_0} dt'' \, (t')^{\nu-\frac{1}{2}}(t'')^{\nu-\frac{1}{2}}\,\delta(t'-t'')
\\ \nonumber
&= \frac{t}{2\nu}\left[ 1 - \left(\frac{t_0}{t}\right)^{2\nu} \right],
\end{align}
where we use a fact that the expectation value of the two noise sources is a $\delta$-function. More precisely,
\begin{equation}
\left\langle \eta(t_1)\eta(t_2) \right\rangle_{\rm S} = \delta (t_1 - t_2).
\end{equation}
The diagrammatic expression of the 2-point function is given by Fig.\ref{scffig2}(e). 

\begin{figure}[b!]
\centering
\includegraphics[width=180mm]{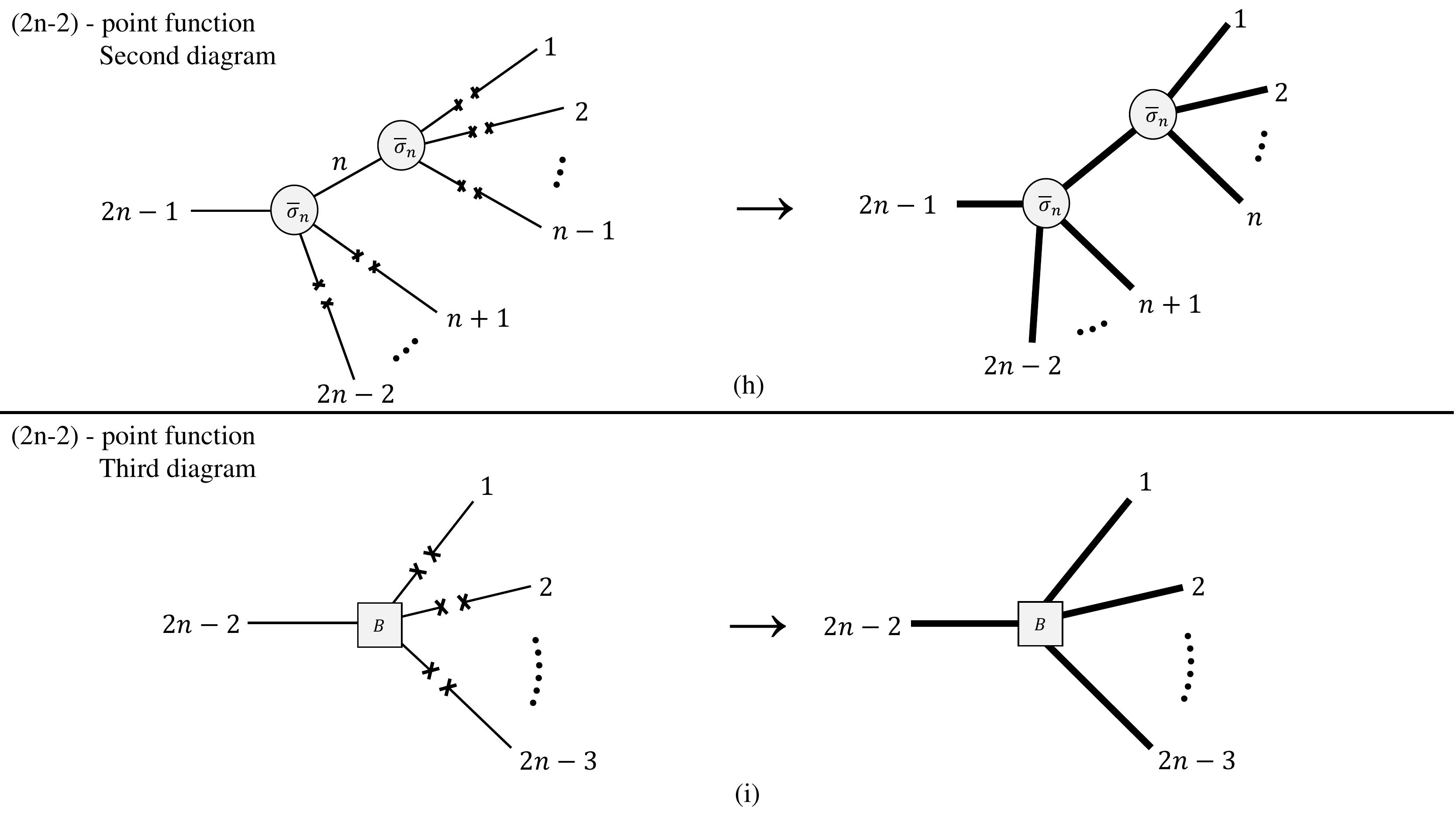}
\caption{Diagrams for the second and the third type $(2n-2)$-point functions.}
\label{scffig3}
\end{figure}

\subsubsection{Stochastic $n$-point function}
Next, we move on to the cases with interactions, $n$-point and $(2n-2)$-point functions. The stochastic $n$-point function at tree level is given by
\begin{align}
\label{stochastic-n-point-function-at-tree-level}
\langle \overbrace{\Phi(t)\cdots\Phi(t)}^{n\ \textrm{fields}} {\rangle}_\textrm{S} &=
n\, \bigg\langle \left[\prod^{n-1}_{j=1}\int^t_{t_0} dt_j\, \frac{t^{\frac{1}{2}-\nu}}{(t_j)^{\frac{1}{2}-\nu}}\,\eta(t_j)\right] \times n\int^t_{t'_0} dt_n\, \frac{t^{\frac{1}{2}-\nu}}{(t_n)^{\frac{1}{2}-\nu}} \left(\bar{\sigma}_n(t_n)^{n(-\frac{1}{2}+\nu)}\right)S_{n-1}(t_n) {\bigg\rangle}_\textrm{S}
\\ \nonumber
&= n^2 \bar{\sigma}_n t^{n(\frac{1}{2}-\nu)} \bigg\langle \left[\prod^{n-1}_{j=1}\int^t_{t_0}dt_j\,(t_j)^{\nu-\frac{1}{2}}\,\eta(t_j)\right] \int^t_{t'_0} dt_n\,(t_n)^{(n+1)(\nu-\frac{1}{2})} S_{n-1}(t_n){\bigg\rangle}_\textrm{S}.
\end{align}
There are $n-1$ times of $\eta$'s in the square brackets and also $n-1$ times of $\eta$'s in $S_{n-1}$. We contract them in such a way that the $n$-point function becomes connected and tree-diagram. In fact, there are $(n-1)!$ ways to do this. Then, we have
\begin{align}
&\bigg\langle \left[\prod^{n-1}_{j=1}\int^t_{t_0}dt_j\,(t_j)^{\nu-\frac{1}{2}}\,\eta(t_j)\right] S_{n-1}(t_n){\bigg\rangle}_\textrm{S}
\\ \nonumber
&= \left[ \prod^{n-1}_{j=1}\int^t_{t_0}dt_j\,(t_j)^{\nu-\frac{1}{2}} \right] \bigg\langle \eta(t_1)\cdots\eta(t_{n-1}) S_{n-1}(t_n){\bigg\rangle}_\textrm{S}
\\ \nonumber
&= \left[ \prod^{n-1}_{j=1}\int^t_{t_0}dt_j\,(t_j)^{\nu-\frac{1}{2}} \prod^{n-1}_{k=1}\int^t_{t_0}dt'_k\,(t'_k)^{\nu-\frac{1}{2}} \right] (t_n)^{(n-1)(\frac{1}{2}-\nu)} \bigg\langle \eta(t_1)\eta(t'_1) \cdots \eta(t_{n-1})\eta(t'_{n-1}) {\bigg\rangle}_\textrm{S}
\\ \nonumber
&= (n-1)!(t_n)^{(1-n)(\nu - \frac{1}{2})} \prod^{n-1}_{j=1}\int^t_{t_0} dt_j \int^{t_n}_{t_0} dt'_j \left(t_j t'_j\right)^{\nu-\frac{1}{2}} \delta(t_j-t'_j).
\end{align}
For the last equality in the above equation, $\eta(t_j)$ needs to be contracted with $\eta(t'_j)$ to make connected diagram only. Therefore, once we plug the above equation into \eqref{stochastic-n-point-function-at-tree-level}, we get
\begin{align}
\langle \overbrace{\Phi(t)\cdots\Phi(t)}^{n\ \textrm{fields}} {\rangle}_\textrm{S}
= n\cdot n! \bar{\sigma}_n t^{n(\frac{1}{2}-\nu)} \int^t_{t'_0} dt_n\,(t_n)^{2\nu-1} \prod^{n-1}_{j=1}\int^t_{t_0} dt_j \int^{t_n}_{t_0} dt'_j \left(t_j t'_j\right)^{\nu-\frac{1}{2}} \delta(t_j-t'_j).
\end{align}
We integrate over $t_j$ first rather than $t'_j$ for every factor in the product, because $(t_0,t_n)\subset(t_0,t)$. Otherwise, we cannot evaluate the $\delta$-function correctly. After the integration, we get the following expression:
\begin{align}
\langle \overbrace{\Phi(t)\cdots\Phi(t)}^{n\ \textrm{fields}} {\rangle}_\textrm{S}
&= n\cdot n! \bar{\sigma}_n t^{n(\frac{1}{2}-\nu)} \int^t_{t'_0} dt_n\,(t_n)^{2\nu-1} \prod^{n-1}_{j=1} \int^{t_n}_{t_0} dt_j \left(t_j\right)^{2\nu-1}
\\ \nonumber
&= n\cdot n! \bar{\sigma}_n t^{n(\frac{1}{2}-\nu)} \int^t_{t'_0} dt_n\,(t_n)^{2\nu-1} \left[ \frac{(t_n)^{2\nu}-(t_0)^{2\nu}}{2\nu} \right]^{n-1}
\\ \nonumber
&= n!\bar{\sigma}_n \left[ \frac{t^{2\nu}-(t_0)^{2\nu}}{2\nu t^{2\nu-1}} \right]^n t^{n(\nu-\frac{1}{2})} - n!\bar{\sigma}_n \left[ \frac{(t'_0)^{2\nu}-(t_0)^{2\nu}}{2\nu t^{2\nu-1}} \right]^n t^{n(\nu-\frac{1}{2})}.
\end{align}
If $t'_0=0$, we finally get
\begin{align}
\langle \overbrace{\Phi(t)\cdots\Phi(t)}^{n\ \textrm{fields}} {\rangle}_\textrm{S}
= n!\bar{\sigma}_n \left[ \frac{t^{2\nu}-(t_0)^{2\nu}}{2\nu t^{2\nu-1}} \right]^n t^{n(\nu-\frac{1}{2})} - n!\bar{\sigma}_n t^{n(\frac{1}{2}-\nu)}\left[ \frac{-(t_0)^{2\nu}}{2\nu} \right]^n.
\end{align}
The diagrammatic expression of the $n$-point function is given by Fig.\ref{scffig2}(f). We note that $t_0'=0$ is the right boundary condition to match the result with the holographic data since the $n$-multiple trace deformation is defined on the conformal boundary $r=0$, which translates $t=0$ via the relation, $r=t$ that the authors suggest in \cite{Oh:2012bx,Jatkar:2013uga,Oh:2013tsa,Oh:2015xva,Moon:2017btx}.

\subsubsection{Stochastic ($2n-2$)-point function}
The stochastic $(2n-2)$-point function is comprised of three different kinds of diagrams. Their evaluations are explained in Appendix B, in detail. Recalling the weak field expansion given in (\ref{Phi-Langevin-sol-final}), we assign the solution of Langevin equation as
\begin{align}
\nonumber
\Phi_2(t) &= \int^t_{t_0}dt'\,\frac{t^{\frac{1}{2}-\nu}}{t'^{\frac{1}{2}-\nu}}\,\eta(t'),
\\ \nonumber
\Phi_n(t) &= n \bar{\sigma}_n t^{\frac{1}{2}-\nu} \int^t_{t'_0}dt'\,(t')^{(n+1)(-\frac{1}{2}+\nu)}S_{n-1}(t'),
\\ \nonumber
\Phi^{(1)}_{2n-2}(t) &= t^{\frac{1}{2}-\nu}\int^t_{t''_0}dt'\ n^2(n-1)\bar{\sigma}_n^2\,(t')^{n(-\frac{1}{2}+\nu)} \left\{ \int^{t'}_{t_0}dt''\,(t'')^{(n+1)(-\frac{1}{2}+\nu)}S_{n-1}(t'') \right\}S_{n-2}(t'),
\\ \nonumber
\Phi^{(2)}_{2n-2}(t) &= t^{\frac{1}{2}-\nu}\int^t_{t''_0}dt'\ (n-1)\left\{ 2\bar{\sigma}_{2n-2}+\left(\frac{\bar{\lambda}_{2n-2}}{n-1}-n^2\bar{\sigma}_n^2 \right)\frac{(t')^{2\nu}}{2\nu} \right\}(t')^{2(n-1)(-\frac{1}{2}+\nu)} S_{2n-3}(t'),
\end{align}
where it is defined that $\Phi_{2n-2}\equiv\sum_{i=1,2}\Phi^{(i)}_{2n-2}$. The ($2n-2$)-point function is comprised of three different diagrams, which are graphically denoted in Fig.\ref{scffig2}(g) and Fig.\ref{scffig3}(h), (i) in order. Therefore, we rewrite the ($2n-2$)-point correlation function as
\begin{equation}
\langle \overbrace{\Phi(t)\cdots\Phi(t)}^{2n-2 \textrm{ fields}}\rangle_\textrm{S}\equiv \sum_{\bf {I}={\bf 1,2,3}}\langle \overbrace{\Phi(t)\cdots\Phi(t)}^{2n-2 \textrm{ fields}}\rangle^{[\bf I]}_\textrm{S},
\end{equation}
and we call the diagrams with $\bf {I}={\bf 1,2,3}$ the first, the second, and the third type diagrams, respectively.

Let us look at the first type of the diagrams.
The first kind is obtained from the expectation value of product of ($2n-4$) times of $\Phi_2(t)$ and square of $\Phi_n(t)$ in such a way that we connect each two $\Phi_n(t)$ by contracting each one of the noise fields in the two $\Phi_n(t)$. The remaining ($2n-4$) noise fields are contracted with those in ($2n-4$) times of $\Phi_2(t)$. Pictorially, it is given in Fig.\ref{scffig2}(g). Its final form is 
\begin{align}
\label{2n-2pointfirsttypediagram}
&
\langle \overbrace{\Phi(t)\cdots\Phi(t)}^{2n-2 \textrm{ fields}}\rangle^{[\bf 1]}_\textrm{S}
=\left\langle\frac{(2n-2)!}{2!(2n-4)!} \Phi_2(t)^{2n-4} \Phi_n(t)^2\right\rangle_S
\\ \nonumber
&= \frac{(2n-2)!}{2!(2n-4)!} n^2 \int^t_{t'_0}dt'\int^t_{t'_0}d\bar{t}' \frac{t^{\frac{1}{2}-\nu} t^{\frac{1}{2}-\nu}}{(t')^{\frac{1}{2}-\nu} (\bar{t}')^{\frac{1}{2}-\nu}} \left[ \bar{\sigma}_n (t')^{n(-\frac{1}{2}+\nu)} \bar{\sigma}_n (\bar{t}')^{n(-\frac{1}{2}+\nu)} \right]
\\ \nonumber
&\hphantom{=} \times \left\langle \left[\prod^{2n-4}_{j=1}\int^t_{t_0}dt_j \frac{t^{\frac{1}{2}-\nu}}{(t_j)^{\frac{1}{2}-\nu}}\,\eta(t_j)\right] S_{n-1}(t') S_{n-1}(\bar{t}') \right\rangle_\textrm{S}
\\ \nonumber
&= \frac{(2n-2)!}{2!}\frac{2}{n}\, n^2(n-1)^2 \bar{\sigma}_n^2\, t^{(n-1)(1-2\nu)}
\left[ \frac{\{t^{2\nu}-(t_0)^{2\nu}\}^{2n-1}-\{(t'_0)^{2\nu}-(t_0)^{2\nu}\}^{2n-1}}{(2n-1)(2\nu)^{2n-1}} \right.
\\ \nonumber
&\hphantom{=\frac{(2n-2)!}{2!}\frac{2}{n}\, n^2(n-1)^2 \bar{\sigma}_n^2\, t^{(n-1)(1-2\nu)}[}
\left. - \frac{\{t^{2\nu}-(t_0)^{2\nu}\}^{n-1}-\{(t'_0)^{2\nu}-(t_0)^{2\nu}\}^{n-1}}{(n-1)(2\nu)^{n-1}} \left(\frac{(t'_0)^{2\nu}-(t_0)^{2\nu}}{2\nu}\right)^n \right].
\end{align}
The second kind is obtained from the expectation value of product of ($2n-3$) times of $\Phi_2(t)$ and $\Phi^{(1)}_{2n-2}(t)$, where all the ($2n-3$) noise fields in $\Phi^{(1)}_{2n-2}(t)$ are contracted with those in ($2n-3$) times of the fields $\Phi_2(t)$. Fig.\ref{scffig3}(h) is this diagram. The computation result is given by
\begin{align}
\label{2n-2pointsecondtypediagram}
&
\langle \overbrace{\Phi(t)\cdots\Phi(t)}^{2n-2\textrm{ fields}} \rangle^{[\bf 2]}_\textrm{S}
=\left\langle \frac{(2n-2)!}{1!(2n-3)!} \Phi_2(t)^{2n-3} \Phi^{(1)}_{2n-2}(t) \right\rangle_S
\\ \nonumber
&= \left\langle (2n-2) \prod^{2n-3}_{j=1} \left[ \int^t_{t_0} d\tilde{t}_j \frac{t^{\frac{1}{2}-\nu}}{(\tilde{t}_j)^{\frac{1}{2}-\nu}} \, \eta(\tilde{t}_j) \right]\right.
\\ \nonumber
&\quad \times\left. \int^t_{t''_0} dt' \Biggr\{ n^2(n-1) \frac{t^{\frac{1}{2}-\nu}}{(t')^{\frac{1}{2}-\nu}} \left[ \bar{\sigma}_n (t')^{(n-1)(-\frac{1}{2}+\nu)} \right] \left[ \int^{t'}_{t'_0} dt'' \, \bar{\sigma}_n (t'')^{(n+1)(-\frac{1}{2}+\nu)}S_{n-1}(t'') \right] S_{n-2}(t') \right\rangle_S
\\ \nonumber
&= (2n-3)!(2n-2)(n-1)n\bar{\sigma}_n^2\,\frac{t^{(2n-2)(\frac{1}{2}-\nu)}}{(2\nu)^{2n-1}} \left\{ \frac{\left[t^{2\nu}-(t_0)^{2\nu}\right]^{2n-1}-\left[(t''_0)^{2\nu}-(t_0)^{2\nu}\right]^{2n-1}}{(2n-1)}\right.
\\ \nonumber
&\hphantom{(2n-3)!(2n-2)(n-1)n\bar{\sigma}_n^2\,\frac{t^{(2n-2)(\frac{1}{2}-\nu)}}{(2\nu)^{2n-1}}}
- \left. \frac{\left[t^{2\nu}-(t_0)^{2\nu}\right]^{n-1}-\left[(t''_0)^{2\nu}-(t_0)^{2\nu}\right]^{n-1}}{n-1}\, \left[(t'_0)^{2\nu}-(t_0)^{2\nu}\right]^n \right\}.
\end{align}
The third diagram is obtained from the expectation value of product of ($2n-3$) times of $\Phi_2(t)$ and $\Phi^{(2)}_{2n-2}(t)$. All the ($2n-3$) noise fields in $\Phi^{(2)}_{2n-2}(t)$ are contracted with those in ($2n-3$) times of the fields $\Phi_2(t)$. This diagram is given in Fig.\ref{scffig3}(i). The final form of the calculation is given by
\begin{align}
\label{2n-2pointthirdtypediagram}
&
\langle \overbrace{\Phi(t)\cdots\Phi(t)}^{2n-2\textrm{ fields}} \rangle^{[\bf 3]}_\textrm{S}
=\left\langle \frac{(2n-2)!}{1!(2n-3)!} \Phi_2(t)^{2n-3} \Phi^{(2)}_{2n-2}(t)\right\rangle_S
\\ \nonumber
&= (2n-2)(n-1) \int^t_{t''_0} dt' \frac{t^{\frac{1}{2}-\nu}}{(t')^{\frac{1}{2}-\nu}} \left\{ 2\bar{\sigma}_{2n-2} + \left(\frac{\bar{\lambda}_{2n-2}}{n-1} - n^2\bar{\sigma}_n^2\right) \frac{t^{2\nu}}{2\nu} \right\} (t')^{2(n-1)(-\frac{1}{2}+\nu)}
\\ \nonumber
&\hphantom{=} \times \left\langle \prod^{2n-3}_{i=1} \int^t_{t_0} d\bar{t}'_i \frac{t^{\frac{1}{2}-\nu}}{(\bar{t}'_i)^{\frac{1}{2}-\nu}}\,\eta(\bar{t}'_i)S_{2n-3}(t') \right\rangle_\textrm{S}
\\ \nonumber
&= (2n-3)!(2n-2)(n-1)\, t^{2(n-1)(\frac{1}{2}-\nu)} \left[ \frac{A}{2n-2} \frac{\left\{t^{2\nu}-(t_0)^{2\nu}\right\}^{2n-2} - \left\{(t''_0)^{2\nu}-(t_0)^{2\nu}\right\}^{2n-2} }{(2\nu)^{2n-2}} \right.
\\ \nonumber
&\hphantom{(2n-3)!(2n-2)(n-1)\, t^{2(n-1)(\frac{1}{2}-\nu)}[}
+ \left. \frac{B}{2n-1} \frac{\left\{t^{2\nu}-(t_0)^{2\nu}\right\}^{2n-1} - \left\{(t''_0)^{2\nu}-(t_0)^{2\nu}\right\}^{2n-1} }{(2\nu)^{2n-1}} \right],
\end{align}
where
\begin{align}
A &= 2\bar{\sigma}_{2n-2} + \left(\frac{\bar{\lambda}_{2n-2}}{n-1} - n^2\bar{\sigma}_n^2\right) \frac{(t_0)^{2\nu}}{2\nu}
= 2\bar{\sigma}_{2n-2} + B\frac{(t_0)^{2\nu}}{2\nu},
\\
B &= \frac{\bar{\lambda}_{2n-2}}{n-1} - n^2\bar{\sigma}_n^2.
\end{align}


\subsection{The relation between multiple trace deformations and stochastic correlation functions}
In this section, we will show that the multiple trace deformations in holographic Wilsonian renormalization group of scalar field theory with mass $M$, self-interaction, $\lambda_{2n-2}$ together with boundary deformation, $\bar \sigma_n$ can be reconstructed by stochastic correlation functions via the relations suggested in (\ref{relation-2point-and-radial-flow-zero-momentum}), (\ref{relation-npoint-ntrace-zero-momentum}) and (\ref{relation-2n-2point-2n-2trace-zero-momentum}) respectively for zero boundary momentum cases. {\it There appear three different constants in stochstic correlation functions, $t_0$, $t_0^\prime$ and $t_0^{\prime\prime}$. They give initial boundary condition. We note that for the reconstruction, we take $1/ \tilde{a}_0 = -t_0^{2\nu}$, $t_0^\prime=0$ and $t_0^{\prime\prime}=0$ in the followings.}

\subsubsection{Double trace deformation and stochastic 2-point function}
We start with the double trace deformation to warm up, given by \eqref{D2-double-trace-deformation-solution} and \eqref{D2-double-trace-deformation-solution-2}. We may rearrange it into
\begin{align}
D^{(2)}(\epsilon)
&= \frac{1}{2}\left( \frac{\frac{1}{2}-\nu}{\epsilon} \right) + \frac{\tilde{a}_0 \nu\epsilon^{2\nu-1}}{1 + \tilde{a}_0\epsilon^{2\nu}},
\end{align}
where $1/ \tilde{a}_0 \equiv \Phi_0 / a_0$.
Let us rewrite the constant, $\tilde a_0$ by $t_0$ by using the suggested relation between $\tilde a_0$ and $t_0$, $1/ \tilde{a}_0 = -t_0^{2\nu}$ to match one side with another. We also replace the variable $\epsilon$ by $t$ by using $\epsilon = t$. These are followed by the relation suggested in \cite{Oh:2012bx,Jatkar:2013uga,Oh:2013tsa,Oh:2015xva,Moon:2017btx} and also given in (\ref{relation-2point-and-radial-flow-zero-momentum}). This is precisely the linear sum of the following quantities we have already obtained in \eqref{Euclidean-action} and \eqref{stochastic-2-point-function-at-tree-level}:
\begin{align}
\frac{1}{2}\frac{\delta^2 S_E}{\delta \Phi^{\hphantom{0} 2}} &= \frac{-\frac{1}{2} + \nu}{t},
\\
\label{stochastic-2-point-function-at-tree-level-section4_5}
\frac{1}{2} \langle \Phi(t)\Phi(t) \rangle _S ^{-1} &= \frac{\nu t^{2\nu -1}}{ t^{2\nu} -  t_0^{2\nu}} = \frac{\tilde{a}_0\nu t^{2\nu-1}}{1+\tilde{a}_0 t^{2\nu}}.
\end{align}
Therefore, we ensure that the relation between the holographic double trace deformation and the stochastic two-point correlation function is given by
\begin{align}
\frac{1}{2} \langle \Phi (r)\Phi (r)  \rangle _H ^{-1} \,\bigg\rvert ^{r=t}
&= -\frac{1}{4}\frac{\delta^2 S_E}{\delta \Phi_0^{\hphantom{0} 2}} + \frac{1}{2} \langle \Phi(t)\Phi(t) \rangle _S ^{-1},
\end{align}
as discussed in \cite{Oh:2012bx,Jatkar:2013uga,Oh:2013tsa,Oh:2015xva,Moon:2017btx}.


\subsubsection{$n$-multiple trace deformation and stochastic $n$-point function}
Now, we look at $n$-point correlation functions. The holographic data for the $n$-multiple trace operator is given by
\begin{equation}
\label{nmultiple1}
n!D^{(n)}(\epsilon) = n!\bar{\sigma}_n \frac{\epsilon^{n\left(\nu-\frac{1}{2}\right)}}{\left( 1 + \tilde{a}_0\epsilon^{2\nu} \right)^n} = \frac{\delta ^n S_B}{\delta\Phi(\epsilon)^n}.
\end{equation}
The Euclidean action will provide
\begin{equation}
\label{nmultiple2}
\frac{1}{2}\frac{\delta^n S_E}{\delta \Phi(\epsilon) ^n} = -n!\bar\sigma_n t^{n\left(\nu-\frac{1}{2}\right)}.
\end{equation}
Finally the stochastic $n$-point function is given by
\begin{align}
\label{nmultiple3}
\langle \overbrace{\Phi(t)\cdots\Phi(t)}^{n\ \textrm{fields}} {\rangle}_\textrm{S}
&= n!\bar\sigma_n \left[ \frac{t^{2\nu}-(t_0)^{2\nu}}{2\nu t^{2\nu-1}} \right]^n t^{n(\nu-\frac{1}{2})} - n!\bar\sigma_n t^{n(\frac{1}{2}-\nu)}\left[ \frac{-(t_0)^{2\nu}}{2\nu} \right]^n
\\ \nonumber
&= \left[ \frac{t^{2\nu}-(t_0)^{2\nu}}{2\nu t^{2\nu -1}} \right]^n n! \left\{ \bar\sigma_n t^{n\left(\nu-\frac{1}{2}\right)} - \bar\sigma_n \cfrac{t^{n\left(\nu-\frac{1}{2}\right)}}{\left[ 1- \cfrac{t^{2\nu}}{(t_0)^{2\nu}}\right]^n} \right\}
\\ \nonumber
&= n!\bar\sigma_n\left[t^{n(\nu-\frac{1}{2})}-\frac{t^{n\left(\nu-\frac{1}{2}\right)}}{1+\tilde{a}_0t^{2\nu}}\right]\left\langle \Phi(t)\Phi(t)\right\rangle^n_{\rm S}.
\end{align}
For the last equality of the above equation, we used \eqref{stochastic-2-point-function-at-tree-level-section4_5} and the boundary condition $\tilde{a}_0 = -\frac{1}{(t_0)^{2\nu}}$.
In sum, with these quantities(\ref{nmultiple1}), (\ref{nmultiple2}) and (\ref{nmultiple3}), we ensure that the relation between the holographic $n$-multiple trace operator and the stochastic $n$-point correlation function is given by
\begin{align}
-\frac{\delta ^n S_B}{\delta\Phi(\epsilon)^n} \bigg\rvert ^{r=t}
&= \langle \overbrace{\Phi(t)\cdots\Phi(t)}^{n\ \textrm{fields}} {\rangle}_\textrm{S}
\, \{\langle \Phi(t)\Phi(t) \rangle _S ^{-1}\}^n + \frac{1}{2}\frac{\delta^n S_E}{\delta \Phi(\epsilon) ^n}.
\end{align}

\subsubsection{$(2n-2)$-multiple trace deformation and stochastic $(2n-2)$-point function}
Finally, we test the relation(\ref{relation-2n-2point-2n-2trace-zero-momentum}) with the stochastic $(2n-2)$-point function at tree level and the radial flow of ($2n-2$)-multiple trace deformation in holographic data. We list ingredients that we need for our computation.
As the first step, we compute a quantity,
\begin{eqnarray}
&\ &\frac{1}{2}\frac{n^2}{(n!)^2}\left\{ \left\langle \Phi(t)^{2} \right\rangle^{-1}_\textrm{S}  \right\}^{2n-1}\left\{ \left\langle \Phi(t)^{n} \right\rangle_\textrm{S}  \right\}^{2} \\ \nonumber
&=&-({t_0})^{2\nu}\frac{n^2\bar\sigma^2 t^{(n-1)(2\nu-1)}}{4\nu } \left\{ 1-\left(\frac{t}{t_0}\right)^{2\nu}-\frac{2}{\left(1-\left(\frac{t}{t_0}\right)^{2\nu}\right)^{n-1}}+\frac{1}{\left(1-\left(\frac{t}{t_0}\right)^{2\nu}\right)^{2n-1}}\right\},
\end{eqnarray}
by using the expressions of stochastic 2- and $n$-point functions.
By using the first and second type diagrams of the stochastic ($2n-2$)-point correlation function, i.e. the expressions(\ref{2n-2pointfirsttypediagram}), (\ref{2n-2pointsecondtypediagram}), we get the following quantity:
\begin{eqnarray}
\label{2n-2-final1}
&\ &
\frac{1}{2}\frac{n^2}{(n!)^2}\left\{ \left\langle \Phi(t)^{2} \right\rangle^{-1}_\textrm{S}  \right\}^{2n-1}\left\{ \left\langle \Phi(t)^{n} \right\rangle_\textrm{S}  \right\}^{2}
-\sum_{I=\bf 1,2}\langle \overbrace{\Phi(t)\cdots\Phi(t)}^{2n-2 \textrm{ fields}}\rangle^{[\bf I]}_\textrm{S} \frac{\left\{ \left\langle \Phi(t)^{2} \right\rangle^{-1}_\textrm{S}  \right\}^{2n-2}}{(2n-2)!}
\\ \nonumber
&=&-({t_0})^{2\nu}\frac{n^2\bar\sigma^2 t^{(n-1)(2\nu-1)}}{4\nu }
\left\{ \frac{1}{\left(1-\left(\frac{t}{t_0}\right)^{2\nu}\right)^{2n-1}}-\frac{1}{\left(1-\left(\frac{t}{t_0}\right)^{2\nu}\right)^{2n-2}} \right. \\ \nonumber
&+&\frac{1-\left(\frac{t}{t_0}\right)^{2\nu}}{2n-1}-\left.\frac{1}{(2n-1)\left(1-\left(\frac{t}{t_0}\right)^{2\nu}\right)^{2n-2}}\right\}.
\end{eqnarray}

As the second step, we compute the variation of Euclidean action with  $(2n-2)$ times, which gives
\begin{align}
\frac{1}{(2n-2)!}\frac{\delta^{2n-2}S_E}{\delta\Phi(t)^{2n-2}}
&= - \left\{ 2\bar{\sigma}_{2n-2}+\left(\frac{\bar{\lambda}_{2n-2}}{n-1}-n^2\sigma_n^2\right)\frac{t^{2\nu}}{2\nu}\right\} t^{2(n-1)(\nu-\frac{1}{2})}
\\ \nonumber
&= - \left( 2\bar{\sigma}_{2n-2}+ B\frac{t^{2\nu}}{2\nu}\right) t^{2(n-1)(\nu-\frac{1}{2})},
\end{align}
where again
\begin{align}
B = \frac{\bar{\lambda}_{2n-2}}{n-1} - n^2\sigma_n^2.
\end{align}
By using the third type diagram of the stochastic ($2n-2$)-point function(\ref{2n-2pointthirdtypediagram}), we also get a quantity being given by
\begin{eqnarray}
\label{2n-2-final2}
&-&\frac{1}{2}\frac{1}{(2n-2)!}\frac{\delta^{2n-2}S_E}{\delta\Phi(t)^{2n-2}}
-\langle \overbrace{\Phi(t)\cdots\Phi(t)}^{2n-2 \textrm{ fields}}\rangle^{[\bf 3]}_\textrm{S} \frac{\left\{ \left\langle \Phi(t)^{2} \right\rangle^{-1}_\textrm{S}  \right\}^{2n-2}}{(2n-2)!}
\\ \nonumber
&=& Y_{2n-2}(t) + t^{(2n-2)(\nu-\frac{1}{2})}\frac{B}{4\nu(2n-1)}\left(-t^{2\nu}_0+t^{2\nu}+\frac{t_0^{2\nu}}{\left(1-\left(\frac{t}{t_0}\right)^{2\nu}\right)^{2n-2}}\right) ,
\end{eqnarray}
where
\begin{align}
\label{2n-2-final3}
Y_{2n-2}(t) \equiv \frac{t^{(2\nu-1)(n-1)}}{   \left(1- \left(\frac{t}{t_0}\right)^{2\nu}\right)^{2(n-1)}   } \bar{\sigma}_{2n-2}. 
\end{align}
We note that we apply our boundary condition for $t^\prime_0=0$ and $t^{\prime\prime}_0=0$ to get these results.

We sum up the equation(\ref{2n-2-final1}) and equation(\ref{2n-2-final2}), and then it becomes the right-hand side of the relation(\ref{relation-2n-2point-2n-2trace-zero-momentum}). Once we replace every $t_0$ by $\tilde a_0$ by using $\tilde a_0=-\frac{1}{t^{2\nu}_0}$, we understand that this reproduce the left-hand side of the relation(\ref{relation-2n-2point-2n-2trace-zero-momentum}) precisely. The exact expression of the left-hand side of the relation(\ref{relation-2n-2point-2n-2trace-zero-momentum}) is given in (\ref{LHS of the 2n-2 relation}).

\section*{Acknowledgement}
J.H.O thanks his W.J. and Y.J. and he thanks God.
This work was supported by the National Research Foundation of Korea(NRF) grant funded by the Korea government(MSIT). (No.2021R1F1A1047930).

\begin{appendices}

\section{Evaluation of the stochastic $(2n-2)$-point correlation functions}
\subsection{The first type diagram}
In this appendix, we will develop stochastic $(2n-2)$-point function to reconstruct holographic data from them. The $(2n-2)$-point function is comprised of three different kinds of diagrams as addressed in Fig.\ref{scffig2}(c) and Fig.\ref{scffig3}(a),(b). The first kind is given by
\begin{align}
\langle \overbrace{\Phi(t)\cdots\Phi(t)}^{2n-2 \textrm{ fields}}\rangle^{[1]}_\textrm{S}
&= \frac{(2n-2)!}{2!(2n-4)!} n^2 \int^t_{t'_0}dt'\int^t_{t'_0}d\bar{t}' \frac{t^{\frac{1}{2}-\nu} t^{\frac{1}{2}-\nu}}{(t')^{\frac{1}{2}-\nu} (\bar{t}')^{\frac{1}{2}-\nu}} \left[ \bar{\sigma}_n (t')^{n(-\frac{1}{2}+\nu)} \bar{\sigma}_n (\bar{t}')^{n(-\frac{1}{2}+\nu)} \right]
\\ \nonumber
&\hphantom{=} \times \left\langle \left[\prod^{2n-4}_{j=1}\int^t_{t_0}dt_j \frac{t^{\frac{1}{2}-\nu}}{(t_j)^{\frac{1}{2}-\nu}}\,\eta(t_j)\right] S_{n-1}(t') S_{n-1}(\bar{t}') \right\rangle_\textrm{S},
\end{align}
where
\begin{align}
S_{n-1}(t') = \prod^{n-1}_{l=1} \int^{t'}_{t_0} \frac{(t')^{\frac{1}{2}-\nu}}{(t_l)^{\frac{1}{2}-\nu}}\,\eta(t_l)dt_l.
\end{align}
In particular, the factors inside the angle bracket is able to be computed as follows:
\begin{align}
&\left\langle S_{2n-4}(t) S_{n-1}(t') S_{n-1}(\bar{t}') \right\rangle_\textrm{S}^\textrm{Tree}
\\ \nonumber
&=\left\langle \prod^{2n-4}_{j=1}\int^t_{t_0}dt_j\frac{t^{\frac{1}{2}-\nu}}{(t_j)^{\frac{1}{2}-\nu}}\,\eta(t_j) \prod^{n-1}_{l=1}\int^{t'}_{t_0}dt'_l\frac{(t')^{\frac{1}{2}-\nu}}{(t'_l)^{\frac{1}{2}-\nu}}\,\eta(t'_l) \prod^{n-1}_{m=1}\int^{\bar{t}'}_{t_0}d\bar{t}'_m\frac{(\bar{t}')^{\frac{1}{2}-\nu}}{(\bar{t}'_m)^{\frac{1}{2}-\nu}}\,\eta(\bar{t}'_m)  \right\rangle_\textrm{S}
\\ \nonumber
&= (n-1)^2 \int^{t'}_{t_0}dt'_{n-1} \frac{(t')^{\frac{1}{2}-\nu}}{(t'_{n-1})^{\frac{1}{2}-\nu}} \int^{\bar{t}'}_{t_0}d\bar{t}'_{n-1} \frac{(\bar{t}')^{\frac{1}{2}-\nu}}{(\bar{t}'_{n-1})^{\frac{1}{2}-\nu}}\,\delta(t'_{n-1}-\bar{t}'_{n-1})
\\ \nonumber
&\hphantom{=} \times \left\langle \prod^{2n-4}_{j=1}\int^t_{t_0}dt_j\frac{t^{\frac{1}{2}-\nu}}{(t_j)^{\frac{1}{2}-\nu}}\,\eta(t_j) \prod^{n-2}_{l=1}\int^{t'}_{t_0}dt'_l\frac{(t')^{\frac{1}{2}-\nu}}{(t'_l)^{\frac{1}{2}-\nu}}\,\eta(t'_l) \prod^{n-2}_{m=1}\int^{\bar{t}'}_{t_0}d\bar{t}'_m\frac{(\bar{t}')^{\frac{1}{2}-\nu}}{(\bar{t}'_m)^{\frac{1}{2}-\nu}}\,\eta(\bar{t}'_m)  \right\rangle_\textrm{S}^\textrm{Tree},
\end{align}
where we contract a factor of $\eta(t'_l)$ with another factor $\eta(\bar{t}'_m)$ to get connected diagram upto tree level. Then, we have
\begin{align}
\label{appendixA_1st_label}
&\left\langle S_{2n-4}(t) S_{n-1}(t') S_{n-1}(\bar{t}') \right\rangle_\textrm{S}^\textrm{Tree}
\\ \nonumber
&= (n-1)^2 (t')^{\frac{1}{2}-\nu}(\bar{t}')^{\frac{1}{2}-\nu} \int^{t'}_{t_0}dt'_{n-1} (t'_{n-1})^{\nu-\frac{1}{2}} \int^{\bar{t}'}_{t_0}d\bar{t}'_{n-1} (\bar{t}'_{n-1})^{\nu-\frac{1}{2}}\,\delta(t'_{n-1}-\bar{t}'_{n-1})
\\ \nonumber
&\hphantom{=} \times (2n-4)!\prod^{n-2}_{j=1} \int^t_{t_0}dt_j \frac{t^{\frac{1}{2}-\nu}}{(t_j)^{\frac{1}{2}-\nu}} \int^{t'}_{t_0}dt'_j \frac{(t')^{\frac{1}{2}-\nu}}{(t'_j)^{\frac{1}{2}-\nu}}\,\delta(t_j-t'_j) \prod^{n-2}_{l=1} \int^t_{t_0} d\bar{t}_l \frac{t^{\frac{1}{2}-\nu}}{(\bar{t}_l)^{\frac{1}{2}-\nu}} \int^{\bar{t}'}_{t_0} d\bar{t}'_l \frac{(\bar{t}')^{\frac{1}{2}-\nu}}{(\bar{t}'_l)^{\frac{1}{2}-\nu}}\,\delta(\bar{t}_l - \bar{t}'_l),
\end{align}
where we contract $\eta(t_j)$ with either $\eta(t'_l)$ or $\eta(\bar{t}'_m)$ to get tree and connected diagram. Now, we perform the $\delta$-function integrations in \eqref{appendixA_1st_label}.
For the $\int^{t'}_{t_0} dt'_{n-1} \int^{\bar{t}'}_{t_0} d\bar{t}'_{n-1}$ integration, we have two different cases for the integration ranges, i.e., $t'>\bar{t}'$ or $t'\leq \bar{t}'$. In fact, the integration depends on step functions in such a way that if $t'>\bar{t}'$, we integrate $\int^{t'}_{t_0} dt'_{n-1}$ first, and otherwise we integrate $\int^{\bar{t}'}_{t_0} d\bar{t}'_{n-1}$ first. We also perform $\int dt'_j$ and $\int d\bar{t}_l$ integrations first rather than performing either $\int dt_j$ or $\int d\bar{t}'_l$ integrations to get correct evaluations of $\delta$-functions. This is because $(t_0,t)\supset (t_0,t')$ and $(t_0,t)\supset (t_0,\bar{t}')$.
Then, \eqref{appendixA_1st_label} becomes
\begin{align}
&\left\langle S_{2n-4}(t) S_{n-1}(t') S_{n-1}(\bar{t}') \right\rangle_\textrm{S}^\textrm{Tree}
\\ \nonumber
&= (n-1)^2(2n-4)!(t')^{\frac{1}{2}-\nu}(\bar{t}')^{\frac{1}{2}-\nu} \left\{ \Theta(t'-\bar{t}') \int^{\bar{t}'}_{t_0} d\bar{t}'_{n-1} (\bar{t}'_{n-1})^{2\nu-1} + \left(1-\Theta(t'-\bar{t}')\right) \int^{t'}_{t_0}dt'_{n-1} (t'_{n-1})^{2\nu-1} \right\}
\\ \nonumber
&\hphantom{=} \times t^{(n-2)(\frac{1}{2}-\nu)}(t')^{(n-2)(\frac{1}{2}-\nu)} \prod^{n-2}_{j=1} \left[ \int^{t'}_{t_0}dt'_j\,(t'_j)^{2\nu-1} \right] t^{(n-2)(\frac{1}{2}-\nu)}(\bar{t}')^{(n-2)(\frac{1}{2}-\nu)} \prod^{n-2}_{l=1} \left[ \int^{\bar{t}'}_{t_0}d\bar{t}'_l\,(\bar{t}'_l)^{2\nu-1} \right].
\end{align}
Finally we get
\begin{align}
&\left\langle S_{2n-4}(t) S_{n-1}(t') S_{n-1}(\bar{t}') \right\rangle_\textrm{S}^\textrm{Tree}
\\ \nonumber
&= t^{(n-2)(\frac{1}{2}-\nu)}(t')^{(n-2)(\frac{1}{2}-\nu)} \left(\frac{(t')^{2\nu}-(t_0)^{2\nu}}{2\nu}\right)^{n-2} t^{(n-2)(\frac{1}{2}-\nu)}(\bar{t}')^{(n-2)(\frac{1}{2}-\nu)} \left(\frac{(\bar{t}')^{2\nu}-(t_0)^{2\nu}}{2\nu}\right)^{n-2}
\\ \nonumber
&\hphantom{=} \times (n-1)^2(2n-4)!(t')^{\frac{1}{2}-\nu}(\bar{t}')^{\frac{1}{2}-\nu} \left\{ \Theta(t'-\bar{t}')\left(\frac{(\bar{t}')^{2\nu}-(t_0)^{2\nu}}{2\nu}\right) - \left(1-\Theta(t-\bar{t}')\right)\left(\frac{(t')^{2\nu}-(t_0)^{2\nu}}{2\nu}\right) \right\}.
\end{align}
In sum, the expression for the first kind diagram of the $(2n-2)$-point function is given by
\begin{align}
\langle \overbrace{\Phi(t)\cdots\Phi(t)}^{2n-2\textrm{ fields}} \rangle^{[1]}_\textrm{S}
&= \frac{(2n-2)!}{2!}\, n^2(n-1)^2 \bar{\sigma}_n^2\, t^{(n-1)(1-2\nu)} \int^t_{t'_0}dt' \int^t_{t'_0}d\bar{t}' (t')^{2\nu-1} (\bar{t}')^{2\nu-1}
\\ \nonumber
&\hphantom{=} \times \left(\frac{(t')^{2\nu}-(t_0)^{2\nu}}{2\nu}\right)^{n-2} \left(\frac{(\bar{t}')^{2\nu}-(t_0)^{2\nu}}{2\nu}\right)^{n-2} \left\{ \Theta(t'-\bar{t}') \frac{(\bar{t}')^{2\nu}-(t_0)^{2\nu}}{2\nu} \right.
\\ \nonumber
&\hphantom{=} \left.+ \left(1-\Theta(t'-\bar{t}')\right)\frac{(t')^{2\nu}-(t_0)^{2\nu}}{2\nu}\right\}
\\ \nonumber
&= A_n(\bar{\sigma}_n) \left[ \int^t_{t'_0}dt' \int^{t'}_{t'_0}d\bar{t}' (t')^{2\nu-1} (\bar{t}')^{2\nu-1} \left(\frac{(t')^{2\nu}-(t_0)^{2\nu}}{2\nu}\right)^{n-2} \left(\frac{(\bar{t}')^{2\nu}-(t_0)^{2\nu}}{2\nu}\right)^{n-1} \right.
\\ \nonumber
&\hphantom{= A_n(\bar{\sigma}_n)[} \left. + \int^t_{t'_0}d\bar{t}' \int^{\bar{t}'}_{t'_0}dt' (t')^{2\nu-1} (\bar{t}')^{2\nu-1} \left(\frac{(t')^{2\nu}-(t_0)^{2\nu}}{2\nu}\right)^{n-1} \left(\frac{(\bar{t}')^{2\nu}-(t_0)^{2\nu}}{2\nu}\right)^{n-2}  \right]
\\ \nonumber
&= \frac{2}{n}A_n(\bar{\sigma}_n) \left[ \frac{\{t^{2\nu}-(t_0)^{2\nu}\}^{2n-1}-\{(t'_0)^{2\nu}-(t_0)^{2\nu}\}^{2n-1}}{(2n-1)(2\nu)^{2n-1}} \right.
\\ \nonumber
&\hphantom{= \frac{2}{n}A_n(\bar{\sigma}_n)[} \left. - \frac{\{t^{2\nu}-(t_0)^{2\nu}\}^{n-1}-\{(t'_0)^{2\nu}-(t_0)^{2\nu}\}^{n-1}}{(n-1)(2\nu)^{n-1}} \left(\frac{(t'_0)^{2\nu}-(t_0)^{2\nu}}{2\nu}\right)^n \right],
\end{align}
where for the first equality we switch integration ranges to respect the step functions, and
\begin{align}
A_n(\bar{\sigma}_n)= \frac{(2n-2)!}{2!}\, n^2(n-1)^2 \bar{\sigma}_n^2\, t^{(n-1)(1-2\nu)}.
\end{align}
Finally, the first term in $(2n-2)$-point function has the following expression
\begin{align}
\label{appendixA1_beforemanip}
\langle \overbrace{\Phi(t)\cdots\Phi(t)}^{2n-2\textrm{ fields}} \rangle^{[1]}_\textrm{S}
&= \frac{(2n-2)!}{2!}\frac{2}{n}\, n^2(n-1)^2 \bar{\sigma}_n^2\, t^{(n-1)(1-2\nu)}
\\ \nonumber
&\hphantom{=} \times \left[ \frac{\{t^{2\nu}-(t_0)^{2\nu}\}^{2n-1}-\{(t'_0)^{2\nu}-(t_0)^{2\nu}\}^{2n-1}}{(2n-1)(2\nu)^{2n-1}} \right.
\\ \nonumber
&\hphantom{=\times[}
\left. - \frac{\{t^{2\nu}-(t_0)^{2\nu}\}^{n-1}-\{(t'_0)^{2\nu}-(t_0)^{2\nu}\}^{n-1}}{(n-1)(2\nu)^{n-1}} \left(\frac{(t'_0)^{2\nu}-(t_0)^{2\nu}}{2\nu}\right)^n \right].
\end{align}
To match this result with the holographic data, we manipulate \eqref{appendixA1_beforemanip} as follows. When we apply the following boundary conditions, $t'_0=0$ and $(t_0)^{2\nu}=-(\tilde{a_0})^{-1}$, the 1st diagram translates into
\begin{align}
&\langle \overbrace{\Phi(t)\cdots\Phi(t)}^{2n-2\textrm{ fields}} \rangle^{[1]}_\textrm{S}\times \frac{1}{(2n-2)!}\left[\left\langle \Phi^2 \right\rangle^{-1}_{\rm S}\right]^{2n-2}
\\ \nonumber
&=\langle \overbrace{\Phi(t)\cdots\Phi(t)}^{2n-2\textrm{ fields}} \rangle^{[1]}_\textrm{S}\times \frac{1}{(2n-2)!}\left( \frac{2\nu\tilde{a_0}t^{2\nu-1}}{1+\tilde{a_0}t^{2\nu}} \right)^{2n-2}
\\ \nonumber
&= \frac{(n-1)^2}{n} \frac{n^2\bar{\sigma^2_n}}{2\nu\tilde{a_0}}\ t^{(n-1)(2\nu-1)} \left[ \frac{(1+\tilde{a_0}t^{2\nu})}{2n-1} - \frac{(1+\tilde{a_0}t^{2\nu})^{-n+1}}{n-1}-\frac{(1+\tilde{a_0}t^{2\nu})^{-2n+2}}{2n-1} + \frac{(1+\tilde{a_0}t^{2\nu})^{-2n+2}}{n-1}
\right]
\end{align}


\subsection{The second type diagram}
The second term of the $(2n-2)$-point correlation function is given by
\begin{align}
&\langle \overbrace{\Phi(t)\cdots\Phi(t)}^{2n-2\textrm{ fields}} \rangle^{[2]}_\textrm{S}
\\ \nonumber
&= \left\langle (2n-2) \prod^{2n-3}_{j=1} \left[ \int^t_{t_0} d\tilde{t}_j \frac{t^{\frac{1}{2}-\nu}}{(\tilde{t}_j)^{\frac{1}{2}-\nu}} \, \eta(\tilde{t}_j) \right]\right.
\\ \nonumber
&\quad \times\left. \int^t_{t''_0} dt' \Biggr\{ n^2(n-1) \frac{t^{\frac{1}{2}-\nu}}{(t')^{\frac{1}{2}-\nu}} \left[ \bar{\sigma}_n (t')^{(n-1)(-\frac{1}{2}+\nu)} \right] \left[ \int^{t'}_{t'_0} dt'' \, \bar{\sigma}_n (t'')^{(n+1)(-\frac{1}{2}+\nu)}S_{n-1}(t'') \right] S_{n-2}(t') \right\rangle_S
\\ \nonumber
&= (2n-2)(n-1)n^2\bar{\sigma}_n^2 t^{(2n-2)(\frac{1}{2}-\nu)} \prod^{2n-3}_{j=1} \int^t_{t_0}d\tilde{t}_j\, \frac{1}{(\tilde{t}_j)^{\frac{1}{2}-\nu}} \int^t_{t_0''} dt' \int^{t'}_{t_0'} dt''\, (t')^{n(\nu-\frac{1}{2})} (t'')^{(n+1)(\nu-\frac{1}{2})}
\\ \nonumber
&\quad \times\left\langle \eta(\tilde{t}_1)\cdots\eta(\tilde{t}_{2n-3})S_{n-1}(t'') S_{n-2}(t') \right\rangle_S,
\end{align}
where
\begin{align*}
&\left\langle \prod^{2n-3}_{j=1} \int^t_{t_0} d\bar{t}_j \frac{1}{(\tilde{t}_j)^{\frac{1}{2}-\nu}}\,\eta(\tilde{t}_j)S_{n-1}(t'')S_{n-2}(t') \right\rangle_{\rm S}
\\
&= \prod^{2n-3}_{j=1} \prod^{n-1}_{l=1} \prod^{n-2}_{m=1} \int^t_{t_0}d\tilde{t}_j \int^{t''}_{t_0} dt''_l \int^{t'}_{t_0} dt'_m \, \frac{1}{(\tilde{t}_j)^{\frac{1}{2}-\nu}} \frac{(t'')^{\frac{1}{2}-\nu}}{(t''_l)^{\frac{1}{2}-\nu}} \frac{(t')^{\frac{1}{2}-\nu}}{(t'_m)^{\frac{1}{2}-\nu}}
\\ \nonumber
&\hphantom{=\prod^{2n-3}_{j=1} \prod^{n-1}_{l=1} \prod^{n-2}_{m=1} \int^t_{t_0}d\tilde{t}_j \int^{t''}_{t_0} dt''_l \int^{t'}_{t_0} dt'_m}
\times \left\langle \eta(\tilde{t}_1)\cdots\eta(\tilde{t}_{2n-3}) \eta(t''_1)\cdots\eta(t''_{n-1}) \eta(t'_1)\cdots\eta(t'_{n-2}) \right\rangle_S
\\
&= \prod^{n-1}_{j=1} \prod^{n-2}_{s=1} \prod^{n-1}_{l=1} \prod^{n-2}_{m=1} \int^t_{t_0}d\tilde{t}_j d\hat{t}_s \int^{t''}_{t_0} dt''_l \int^{t'}_{t_0} dt'_m \, \frac{1}{(\tilde{t}_j)^{\frac{1}{2}-\nu}} \frac{(t'')^{\frac{1}{2}-\nu}}{(t''_l)^{\frac{1}{2}-\nu}} \frac{(t')^{\frac{1}{2}-\nu}}{(t'_m)^{\frac{1}{2}-\nu}} (2n-3)! \prod^{n-1}_{r=1}\prod^{n-2}_{r'=1}\delta(\tilde{t}_r - t''_r)\delta(\hat{t}_{r'} - t'_{r'})
\\
&= (2n-3)! \prod^{n-1}_{l=1} \prod^{n-2}_{m=1} \int^{t''}_{t_0} dt''_l \int^{t'}_{t_0} dt'_m \frac{(t'')^{\frac{1}{2}-\nu}}{(t''_l)^{1-2\nu}} \frac{(t')^{\frac{1}{2}-\nu}}{(t'_m)^{1-2\nu}}
\\
&= \frac{(2n-3)!}{(2\nu)^{2n-3}} (t'')^{(n-1)(\frac{1}{2}-\nu)} \left[ (t'')^{2\nu} - (t_0)^{2\nu} \right]^{n-1} (t')^{(n-2)(\frac{1}{2}-\nu)} \left[ (t')^{2\nu} - (t_0)^{2\nu} \right]^{n-2}.
\end{align*}
We plug this factor in the 2nd kind diagram, and get
\begin{align}
\langle \overbrace{\Phi(t)\cdots\Phi(t)}^{2n-2\textrm{ fields}} \rangle^{[2]}_\textrm{S}
&= (2n-3)!(2n-2)(n-1)n^2\bar{\sigma}_n^2\,\frac{t^{(2n-2)(\frac{1}{2}-\nu)}}{(2\nu)^{2n-3}}
\\ \nonumber
&\hphantom{=}
\times \int^t_{t''_0} dt' \left\{ \int^{t'}_{t'_0} dt''\, (t'')^{2\nu-1}\left[(t'')^{2\nu} - (t''_0)^{2\nu} \right]^{n-1} \right\} (t')^{2\nu-1}\left[(t')^{2\nu} - (t'_0)^{2\nu} \right]^{n-2}
\\ \nonumber
&= (2n-3)!(2n-2)(n-1)n\bar{\sigma}_n^2\,\frac{t^{(2n-2)(\frac{1}{2}-\nu)}}{(2\nu)^{2n-1}} \left\{ \frac{\left[t^{2\nu}-(t_0)^{2\nu}\right]^{2n-1}-\left[(t''_0)^{2\nu}-(t_0)^{2\nu}\right]^{2n-1}}{(2n-1)}\right.
\\ \nonumber
&- \left. \frac{\left[t^{2\nu}-(t_0)^{2\nu}\right]^{n-1}-\left[(t''_0)^{2\nu}-(t_0)^{2\nu}\right]^{n-1}}{n-1}\, \left[(t'_0)^{2\nu}-(t_0)^{2\nu}\right]^n \right\}.
\end{align}
We apply the boundary conditions $t''_0 = t'_0 = 0$ and $(t_0)^{2\nu} = -(\tilde{a}_0)^{-1}$ to match the result with the holographic data. The 2nd diagram translates into
\begin{align}
&\langle \overbrace{\Phi(t)\cdots\Phi(t)}^{2n-2\textrm{ fields}} \rangle^{[2]}_\textrm{S} \times \frac{1}{(2n-2)!} \left[ \langle \Phi^2 \rangle^{-1}_\textrm{S} \right]^{2n-2}
\\ \nonumber
&= \langle \overbrace{\Phi(t)\cdots\Phi(t)}^{2n-2\textrm{ fields}} \rangle^{[2]}_\textrm{S} \times \frac{1}{(2n-2)!}\left( \frac{2\nu\tilde{a}_0 t^{2\nu-1}}{1 + \tilde{a}_0 t^{2\nu}} \right)^{2n-2}
\\ \nonumber
&= \frac{n-1}{n} \frac{n^2\bar{\sigma}_n^2}{2\nu\tilde{a}_0} t^{(n-1)(2\nu-1)} \left[ \frac{1+\tilde{a}_0 t^{2\nu}}{2n-1} - \frac{(1+\tilde{a}_0 t^{2\nu})^{-n+1}}{n-1} - \frac{(1+\tilde{a}_0 t^{2\nu})^{-2n+2}}{2n-1} + \frac{(1+\tilde{a}_0 t^{2\nu})^{-2n+2}}{n-1} \right]
\end{align}

\subsection{The third type diagram}
The third term of the (2n-2)-point correlation function is given by
\begin{align}
\label{appendixA1_3rd_diagram}
\langle \overbrace{\Phi(t)\cdots\Phi(t)}^{2n-2\textrm{ fields}} \rangle^{[3]}_\textrm{S}
&= (2n-2)(n-1) \int^t_{t''_0} dt' \frac{t^{\frac{1}{2}-\nu}}{(t')^{\frac{1}{2}-\nu}} \left\{ 2\bar{\sigma}_{2n-2} + \left(\frac{\bar{\lambda}_{2n-2}}{n-1} - n^2\bar{\sigma}_n^2\right) \frac{t^{2\nu}}{2\nu} \right\}
\\ \nonumber
&\hphantom{=(2n-2)(n-1) \int^t_{t''_0}}
\times (t')^{2(n-1)(-\frac{1}{2}+\nu)} \left\langle \prod^{2n-3}_{i=1} \int^t_{t_0} d\bar{t}'_i \frac{t^{\frac{1}{2}-\nu}}{(\bar{t}'_i)^{\frac{1}{2}-\nu}}\,\eta(\bar{t}'_i)S_{2n-3}(t') \right\rangle_\textrm{S}.
\end{align}
We compute the last factors inside the angle bracket in \eqref{appendixA1_3rd_diagram} as follows:
\begin{align}
&\left\langle \prod^{2n-3}_{i=1} \int^t_{t_0} d\bar{t}'_i \frac{t^{\frac{1}{2}-\nu}}{(\bar{t}'_i)^{\frac{1}{2}-\nu}}\,\eta(\bar{t}'_i)S_{2n-3}(t') \right\rangle_\textrm{S}
\\ \nonumber
&= \left\langle \prod^{2n-3}_{i=1} \int^t_{t_0} d\bar{t}'_i \frac{t^{\frac{1}{2}-\nu}}{(\bar{t}'_i)^{\frac{1}{2}-\nu}}\,\eta(\bar{t}'_i) \prod^{2n-3}_{j=1} \int^{t'}_{t_0} d\bar{t}_j \frac{(t')^{\frac{1}{2}-\nu}}{(\bar{t}_j)^{\frac{1}{2}-\nu}}\,\eta(\bar{t}_j) \right\rangle_\textrm{S}
\\ \nonumber
&= (2n-3)!\, t^{(2n-3)(\frac{1}{2}-\nu)} (t')^{(2n-3)(\frac{1}{2}-\nu)} \left[ \Theta(t-t')\left( \frac{(t')^{2\nu}-(t_0)^{2\nu}}{2\nu} \right)^{2n-3} + \left( 1-\Theta(t-t') \right) \left( \frac{t^{2\nu}-(t_0)^{2\nu}}{2\nu} \right)^{2n-3} \right].
\end{align}
Because $t\geq t'$, the step function $\Theta(t-t')$ is always equal to $1$. Therefore we have
\begin{align}
&\langle \overbrace{\Phi(t)\cdots\Phi(t)}^{2n-2\textrm{ fields}} \rangle^{[3]}_\textrm{S}
\\ \nonumber
&= (2n-3)!(2n-2)(n-1)\, t^{2(n-1)(\frac{1}{2}-\nu)} \int^t_{t''_0} dt' \left\{ 2\bar{\sigma}_{2n-2} + \left(\frac{\bar{\lambda}_{2n-2}}{n-1} - n^2\bar{\sigma}_n^2\right) \frac{t^{2\nu}}{2\nu} \right\}
\\ \nonumber
&\hphantom{= (2n-3)!(2n-2)(n-1)\, t^{2(n-1)(\frac{1}{2}-\nu)} \int^t_{t''_0} dt'\{ }
\times \left( \frac{(t')^{2\nu}-(t_0)^{2\nu}}{2\nu} \right)^{2n-3} (t')^{2\nu-1}
\\ \nonumber
&= (2n-3)!(2n-2)(n-1)\, t^{2(n-1)(\frac{1}{2}-\nu)} \left[ \frac{A}{2n-2} \frac{\left\{t^{2\nu}-(t_0)^{2\nu}\right\}^{2n-2} - \left\{(t''_0)^{2\nu}-(t_0)^{2\nu}\right\}^{2n-2} }{(2\nu)^{2n-2}} \right.
\\ \nonumber
&\hphantom{(2n-3)!(2n-2)(n-1)\, t^{2(n-1)(\frac{1}{2}-\nu)}[}
+ \left. \frac{B}{2n-1} \frac{\left\{t^{2\nu}-(t_0)^{2\nu}\right\}^{2n-1} - \left\{(t''_0)^{2\nu}-(t_0)^{2\nu}\right\}^{2n-1} }{(2\nu)^{2n-1}} \right],
\end{align}
where
\begin{align}
A &= 2\bar{\sigma}_{2n-2} + \left(\frac{\bar{\lambda}_{2n-2}}{n-1} - n^2\bar{\sigma}_n^2\right) \frac{(t_0)^{2\nu}}{2\nu}
= 2\bar{\sigma}_{2n-2} + B\frac{(t_0)^{2\nu}}{2\nu},
\\
B &= \frac{\bar{\lambda}_{2n-2}}{n-1} - n^2\bar{\sigma}_n^2.
\end{align}
We apply the boundary conditions $t''_0 = t'_0 = 0$ and $(t_0)^{2\nu} = -(\tilde{a}_0)^{-1}$ to match the result with the holographic data. The 3rd diagram translates into
\begin{align}
&\langle \overbrace{\Phi(t)\cdots\Phi(t)}^{2n-2\textrm{ fields}} \rangle^{[3]}_\textrm{S} \times \frac{1}{(2n-2)!} \left[ \langle \Phi^2 \rangle^{-1}_\textrm{S} \right]^{2n-2}
\\ \nonumber
&= \langle \overbrace{\Phi(t)\cdots\Phi(t)}^{2n-2\textrm{ fields}} \rangle^{[3]}_\textrm{S} \times \frac{1}{(2n-2)!}\left( \frac{2\nu\tilde{a}_0 t^{2\nu-1}}{1 + \tilde{a}_0 t^{2\nu}} \right)^{2n-2}
\\ \nonumber
&= (n-1)t^{(n-1)(2\nu-1)} \left[ \frac{A}{2n-2} - \frac{A(1+\tilde{a}_0 t^{2\nu})^{-2n+2}}{2n-2} - \frac{B(1+\tilde{a}_0 t^{2\nu})}{2\nu\tilde{a_0}(2n-1)} + \frac{B(1+\tilde{a}_0 t^{2\nu})^{-2n+2}}{2\nu\tilde{a_0}(2n-1)} \right]
\\ \nonumber
&= \bar{\sigma}_{2n-2}\ t^{(n-1)(2\nu-1)} - Y_{2n-2}(t) - t^{(n-1)(2\nu-1)}\frac{B}{2\nu}\left[\frac{1}{2(2n-1)}\left( \frac{1}{\tilde{a_0}} +t^{2\nu} -\frac{1}{\tilde{a_0}(1+\tilde{a_0}t^{2\nu})^{2n-2}} \right) -\frac{t^{2\nu}}{2} \right],
\end{align}
where
\begin{equation}
Y_{2n-2}(t) = \frac{t^{(n-1)(2\nu-1)}}{(1+\tilde{a}_0 t^{2\nu})^{2n-2}}\ \bar{\sigma}_{2n-2}.
\end{equation}
\end{appendices}

\begin{appendices}
\section{Evaluation of the integral solution of $D^{(2n-2)}(\epsilon)$}
The integral solution of $D^{(2n-2)}(\epsilon)$ is given by
\begin{equation}
\label{integral-solution-d2}
D^{(2n-2)}(\epsilon) = \Phi^{-2(n-1)}(\epsilon)\left[C_{2n-2}+\int^\epsilon_0 d\bar{\epsilon}\, \Phi(\bar{\epsilon})^{2(n-1)}\bar{\epsilon}^{(n-2)(d-1)-2}\left\{ \frac{\bar{\lambda}_{2n-2}}{2(n-1)} - \frac{n^2\sigma_n^2}{2}\left(1+\tilde{a}_0\bar{\epsilon}^{2\nu}\right)^{-2n} \right\} \right],
\end{equation}
where the field $\Phi$ is given by
\begin{equation}
\label{phi}
\Phi(\epsilon) = \Phi_0\epsilon^{\frac{1}{2}-\nu}+a_0\epsilon^{\frac{1}{2}+\nu} = \Phi_0\epsilon^{\frac{1}{2}-\nu} \left( 1+\tilde{a}_0 \epsilon^{2\nu} \right).
\end{equation}
We plug the equation (\ref{phi}) into the equation (\ref{integral-solution-d2}) and get
\begin{align}
\label{integral-solution-d2n-2}
D^{(2n-2)}(\epsilon)&= \Phi_0^{-2(n-1)}\epsilon^{-2(n-1)(\frac{1}{2}-\nu)} \left(1+\tilde{a}_0\epsilon^{2\nu}\right)^{-2(n-1)} \left[C_{2n-2} \vphantom{\frac{\bar{\lambda}_{2n-2}}{(n-1)}}\right.
\\ \nonumber
&+\left.\frac{1}{2}\int^\epsilon_0 d\bar{\epsilon}\, \Phi_0^{2(n-1)}\bar{\epsilon}^{2(n-1)(\frac{1}{2}-\nu)} \left(1+\tilde{a}_0\bar{\epsilon}^{2\nu}\right)^{2(n-1)} \bar{\epsilon}^{(n-2)(d-1)-2} \left\{ \frac{\bar{\lambda}}{(n-1)} - n^2\sigma_n^2\left(1+\tilde{a}_0\bar{\epsilon}^{2\nu}\right)^{-2n} \right\} \right].
\end{align}
We note that the coefficient can be determined by the relation
\begin{equation}
C_{2n-2}\Phi_0^{-2(n-1)} \equiv \bar{\sigma}_{2n-2},
\end{equation}
where $\bar{\sigma}_{2n-2}$ is the ($2n-2$)-multiple trace coupling.
Now we evaluate $\int^{\epsilon}_0 d\bar{\epsilon}$ integration in the second line of the equation (\ref{integral-solution-d2n-2}) with the determined coefficient. The result is given by
\begin{align}
\nonumber
&\epsilon^{(2\nu-1)(n-1)} \left(1+\tilde{a}_0\epsilon^{2\nu}\right)^{-2(n-1)} \left[ \bar{\sigma}_{2n-2} + \frac{\bar{\lambda}_{2n-2}}{2(n-1)} \left.\frac{\left(1+\tilde{a}_0\bar{\epsilon}^{2\nu}\right)^{2(n-1)}}{(2n-1)2\nu\tilde{a}_0} \right\rvert^\epsilon_0 \right.
+\left.\left.\frac{n^2\sigma_n^2}{2}\frac{\left(1+\tilde{a}_0\bar{\epsilon}^{2\nu}\right)^{-1}}{2\nu\tilde{a}_0}\right\rvert^\epsilon_0\ \right]
\\ \nonumber
&= \frac{\epsilon^{(2\nu-1)(n-1)}}{\left(1+\tilde{a}_0\bar{\epsilon}^{2\nu}\right)^{2(n-1)}}\, \bar{\sigma}_{2n-2} + \frac{\epsilon^{(2\nu-1)(n-1)}}{\left(1+\tilde{a}_0\bar{\epsilon}^{2\nu}\right)^{2(n-1)}} \left[ \frac{\bar{\lambda}_{2n-2}}{2(n-1)}\frac{1}{(2n-1)2\nu\tilde{a}_0}\left\{\left(1+\tilde{a}_0\epsilon^{2\nu}\right)^{2n-1}-1\right\}  \right.
\\ \nonumber
&\left. + \frac{n^2\sigma_n^2}{2}\frac{1}{2\nu\tilde{a}_0} \left\{\left(1+\tilde{a}_0\epsilon^{2\nu}\right)^{-1}-1\right\} \right].
\end{align}
Finally, the solution of $D^{(2n-2)}(\epsilon)$ is given by
\begin{align}
\nonumber
D^{(2n-2)}&= Y_{2n-2}(\epsilon) + \epsilon^{(2\nu-1)(n-1)} \left[ \frac{B+n^2\sigma_n^2}{(2n-1)4\nu\tilde{a}_0} \left\{(1+\tilde{a}_0\epsilon^{2\nu})-\frac{1}{(1+\tilde{a}_0\epsilon^{2\nu})^{2n-2}}\right\} \right.
\\ \nonumber
&\left. +\frac{n^2\sigma_n^2}{4\nu\tilde{a}_0} \left\{\frac{1}{(1+\tilde{a}_0\epsilon^{2\nu})^{2n-1}} - \frac{1}{(1+\tilde{a}_0\epsilon^{2\nu})^{2n-2}}\right\} \right],
\end{align}
where
\begin{equation}
Y_{2n-2}(\epsilon) \equiv \frac{\epsilon^{(2\nu-1)(n-1)}}{\left(1+\tilde{a}_0 \epsilon^{2\nu}\right)^{2(n-1)}}\, \bar{\sigma}_{2n-2},
\end{equation}
and
\begin{equation}
    B=\frac{\bar{\lambda}_{2n-2}}{n-1}-n^2\bar{\sigma}^2.
\end{equation}
\end{appendices}


\begin{thebibliography}{9}

\bibitem{Peskin11}
Michael E. Peskin, Daniel V. Schroeder,  ``An Introduction To Quantum Field Theory''


\bibitem{Parisi:1980ys} 
  G.~Parisi and Y.~s.~Wu,
  Sci.\ Sin.\  {\bf 24}, 483 (1981).

\bibitem{Damgaard:1987rr} 
  P.~H.~Damgaard and H.~Huffel,
  Phys.\ Rept.\  {\bf 152}, 227 (1987).
  doi:10.1016/0370-1573(87)90144-X


\bibitem{Lifschytz:2000bj}
G.~Lifschytz and V.~Periwal,
JHEP \textbf{04}, 026 (2000)
doi:10.1088/1126-6708/2000/04/026
[arXiv:hep-th/0003179 [hep-th]].


\bibitem{Mansi:2009mz}
D.~S.~Mansi, A.~Mauri and A.~C.~Petkou,
Phys. Lett. B \textbf{685}, 215-221 (2010)
doi:10.1016/j.physletb.2010.01.033
[arXiv:0912.2105 [hep-th]].



\bibitem{Oh:2012bx} 
  J.~H.~Oh and D.~P.~Jatkar,
  JHEP {\bf 1211}, 144 (2012)
  doi:10.1007/JHEP11(2012)144
  [arXiv:1209.2242 [hep-th]].

\bibitem{Jatkar:2013uga} 
  D.~P.~Jatkar and J.~H.~Oh,
  JHEP {\bf 1310}, 170 (2013)
  doi:10.1007/JHEP10(2013)170
  [arXiv:1305.2008 [hep-th]].

\bibitem{Oh:2013tsa} 
  J.~H.~Oh,
  Int.\ J.\ Mod.\ Phys.\ A {\bf 29}, 1450082 (2014)
  doi:10.1142/S0217751X14500821
  [arXiv:1310.0588 [hep-th]].

\bibitem{Oh:2015xva} 
  J.~H.~Oh,
  Phys.\ Rev.\ D {\bf 94}, no. 10, 105020 (2016)
  doi:10.1103/PhysRevD.94.105020
  [arXiv:1504.03046 [hep-th]].

\bibitem{Moon:2017btx} 
  S.~p.~Moon,
  Int.\ J.\ Mod.\ Phys.\ A {\bf 33}, no. 16, 1850091 (2018)
  doi:10.1142/S0217751X18500914
  [arXiv:1702.00117 [hep-th]].
  
\bibitem{Heemskerk:2010hk}  
  I.~Heemskerk and J.~Polchinski,
  JHEP {\bf 1106}, 031 (2011)
  doi:10.1007/JHEP06(2011)031
  [arXiv:1010.1264 [hep-th]].
  
\bibitem{Faulkner:2010jy} 
  T.~Faulkner, H.~Liu and M.~Rangamani,
  JHEP {\bf 1108}, 051 (2011)
  doi:10.1007/JHEP08(2011)051
  [arXiv:1010.4036 [hep-th]].
  
\bibitem{Witten:2001ua} 
  E.~Witten,
  hep-th/0112258.

\bibitem{Aharony:2015afa} 
  O.~Aharony, G.~Gur-Ari and N.~Klinghoffer,
  JHEP {\bf 1505}, 031 (2015)
  doi:10.1007/JHEP05(2015)031
  [arXiv:1501.06664 [hep-th]].

\bibitem{Mann:2011hg}
R.~B.~Mann and R.~McNees,
JHEP \textbf{10}, 129 (2011)
doi:10.1007/JHEP10(2011)129
[arXiv:1107.5792 [hep-th]].

\bibitem{Chemissany:2012du}
W.~Chemissany, D.~Geissbuhler, J.~Hartong and B.~Rollier,
Class. Quant. Grav. \textbf{29}, 235017 (2012)
doi:10.1088/0264-9381/29/23/235017
[arXiv:1205.5777 [hep-th]].

\bibitem{Kim:2021rik}
G.~Kim and J.~H.~Oh,
J. Korean Phys. Soc. \textbf{80}, no.1, 30-36 (2022)
doi:10.1007/s40042-021-00357-y
[arXiv:2111.07600 [hep-th]].

\bibitem{deHaro:2006ymc}
S.~de Haro, I.~Papadimitriou and A.~C.~Petkou,
Phys. Rev. Lett. \textbf{98}, 231601 (2007)
doi:10.1103/PhysRevLett.98.231601
[arXiv:hep-th/0611315 [hep-th]].

\bibitem{Oh:2021bxx}
J.~H.~Oh,
J. Korean Phys. Soc. \textbf{79}, no.10, 903-917 (2021)
doi:10.1007/s40042-021-00320-x
[arXiv:2110.05013 [hep-th]].

\bibitem{Jatkar:2012mm}
D.~P.~Jatkar and J.~H.~Oh,
JHEP \textbf{08}, 077 (2012)
doi:10.1007/JHEP08(2012)077
[arXiv:1203.2106 [hep-th]].

\bibitem{Oh:2014nfa}
J.~H.~Oh,
Int. J. Mod. Phys. A \textbf{30}, no.17, 1550098 (2015)
doi:10.1142/S0217751X15500980
[arXiv:1411.6356 [hep-th]].

\bibitem{Oh:2020zvm}
J.~H.~Oh,
JHEP \textbf{11}, 100 (2020)
doi:10.1007/JHEP11(2020)100
[arXiv:2005.08521 [hep-th]].
\end{thebibliography}
\end{document}